\documentclass[twocolumn,pra,showpacs,amsmath]{revtex4}
\usepackage{epsfig}

\def\be{\begin{equation}}
\def\ee{\end{equation}}
\def\e#1{\label{#1}\end{equation}}
\def\bea{\begin{eqnarray}}
\def\eea{\end{eqnarray}}
\def\ea#1{\label{#1}\end{eqnarray}}
\def\rqn#1{(\ref{#1})}
\def\bes#1{\begin{subequations}\label{#1}}
\def\ese{\end{subequations}}

\begin{document}

\title{Effects of decoherence and errors on Bell-inequality
violation}
\author{Abraham G. Kofman}
\affiliation{Department of Electrical Engineering, University of
California, Riverside, California 92521}
\date{\today}

\begin{abstract}
 We study optimal conditions for violation of the
Clauser-Horne-Shimony-Holt form of the Bell inequality in the
presence of decoherence and measurement errors.
 We obtain all detector configurations providing the maximal Bell
inequality violation for a general (pure or mixed) state.
 We consider local decoherence which includes energy relaxation at
the zero temperature and arbitrary dephasing.
 Conditions for the maximal Bell-inequality violation in the presence
of decoherence are analyzed both analytically and numerically for the
general case and for a number of important special cases.
 Combined effects of measurement errors and decoherence are also
discussed.
 \end{abstract}

\pacs{03.65.Ud, 03.65.Yz, 85.25.Cp}
 \maketitle

\section{Introduction}

Entanglement, i.e., quantum nonlocal correlations between physical
systems, is not only a basic feature of quantum behavior
\cite{sch35,EPR}, but also an important resource for quantum
computation and quantum information \cite{nie00}.
 Decoherence, i.e., loss of coherence of states of quantum systems
due to the interaction with the environment, is one of the major
stumbling blocks for quantum computation \cite{nie00}.
 Therefore recently there has been a surge of interest in effects
of decoherence on entanglement
\cite{jak03,jak04,tyu04,tol05,sli05,Nori-06,jam06,ged06,san06,ban07,
kof}.

One of the most striking manifestations of the nonclassical nature of
entanglement is violation of the Bell inequality (BI) \cite{Bell}.
 Effects of decoherence on the BI violation have attracted a
significant interest recently
\cite{jak03,jak04,sli05,jam06,ged06,kof,sam03,vel03,kof08}.
 Decoherence transforms a pure entangled state into a mixed state,
decreasing thus entanglement and the Bell inequality violation.
 While any pure (completely coherent) entangled state can be used
for violation of the BI \cite{cap73}, there are mixed (partially
incoherent) entangled states which cannot violate the BI
\cite{wer89}.
 In fact, the ratio of the volume of the entangled states violating
the BI to the volume of the entangled states obeying the BI is small.
 In particular, in the Hilbert-Schmidt metric this ratio equals
$0.01085$ \cite{kof}.
 This suggests that observation of the Bell inequality violation is
a more difficult task than observation of entanglement.
 Indeed, in the presence of decoherence the duration of the
Bell-inequality violation is generally significantly shorter than the
entanglement survival time \cite{kof}.

Until now most experiments on the BI violation have been performed
with photons \cite{asp81,Weihs,asp02,alt05}.
 However, recently there has been an increasing interest in testing
the BI for various material systems, where decoherence is usually an
important factor.
 In particular, experiments with ions in traps \cite{Rowe}, an
atom-photon system \cite{moe04}, and single neutrons
\cite{Hasegawa03} were performed.
 There are also various theoretical proposals related to the BI
violation in solid-state systems
\cite{sam03,vel03,Ionicioiu,Samuelsson-05,Trauzettel-06, Nori-05}.
 Experiments on testing the BI in superconducting Josephson phase
qubits \cite{mar02} are currently underway \cite{ans06}.

Optimal experimental conditions for observation of the BI violation
in superconducting phase qubits were considered in Ref.
\onlinecite{kof08}.
 Both the ideal case and effects of various nonidealities, such as
measurement errors and crosstalk \cite{mcd05,kof07}, were analyzed in
detail, while decoherence was discussed briefly.
 In the recent paper \cite{kof} the relation between
entanglement and the BI violation was considered.
 In particular, the survival times of entanglement and the BI
violation in the presence of local decoherence were studied, in the
frame of a decoherence model more general than in previous
publications \cite{jak03,jak04,tyu04,tol05,sli05,Nori-06,
jam06,ged06,san06,ban07,sam03,vel03}, the focus being on the class of
``odd'' two-qubit states relevant for experiments with
superconducting phase qubits \cite{mcd05,ste06}.
 Note, however, that optimal detector configurations providing the
maximal BI violation were not discussed previously, except for the
case of pure dephasing \cite{sam03} (see also brief remarks in Refs.
\onlinecite{kof} and \onlinecite{kof08}).
 The knowledge of all optimal configurations is important for
planning experiments, since some detector configurations can be
easier to realize than others \cite{sam03}.

In the present paper, we provide a comprehensive discussion of
effects of local decoherence on the BI violation.
 We obtain both optimal states and all detector configurations which
yield the maximal BI violation.
 Our decoherence model includes energy relaxation at the zero
temperature (known also as spontaneous decay or amplitude damping)
and dephasing (phase damping).
 We analyze analytically and numerically the general case and a
number of important special cases.
 In particular, we study the experimentally relevant classes of the
``even'' and ''odd'' states.
 We also discuss the combined effect of decoherence and local
errors, basing on the treatment of errors in Ref. \onlinecite{kof08}.

The paper is organized as follows.
 In Sec. \ref{II} we discuss the BI and properties of maximally
entangled states.
 In Sec. \ref{III} we obtain all optimal configurations of the
detectors which maximize the BI violation for any given (pure or
mixed) state.
 Section \ref{IV} is devoted to effects of local (independent)
decoherence of the qubits on the BI violation.
 In Sec. \ref{V} we consider combined effects of errors and
decoherence.
 Section \ref{VI} provides the concluding remarks.

\section{Preliminaries}
 \label{II}

In this section we introduce the notation and review properties of
the BI and maximally entangled states.
 Here and below till Sec. \ref{V} we assume the absence of
measurement errors.

\subsection{The Bell inequality}
 \label{IIA}

We consider a pair of qubits, i.e., two-level systems $a$ and $b$.
 Each qubit has the states $|0\rangle$ and $|1\rangle$.
 Measurements of two qubits satisfy the Bell
inequality, provided that a realistic (classical) theory holds and
there is no communication between the qubits \cite{Bell}.
 We consider the Clauser-Horne-Shimony-Holt version of the BI
\cite{CHSH,bel71}
 \be
-2\le S\le2,
 \e{3}
where
 \be
S=E(\vec{a},\vec{b})-E(\vec{a},\vec{b}')+E(\vec{a}',\vec{b})+
E(\vec{a}',\vec{b}').
 \e{4}
Here $\vec{a}$ and $\vec{a}'$ ($\vec{b}$ and $\vec{b}'$) are the
unit radius-vectors on the Bloch sphere along the measurement
(detector) axes for qubit $a$ ($b$) and $E(\vec{a},\vec{b})$ is a
correlator of the measurement results for the qubits $a$ and $b$ in
configurations $\vec{a}$ and $\vec{b}$, respectively,
 \be
E(\vec{a},\vec{b})=p_{11}(\vec{a},\vec{b})+p_{00}(\vec{a},\vec{b})-
p_{10}(\vec{a},\vec{b})-p_{01}(\vec{a},\vec{b}),
 \e{5}
where $p_{ij}(\vec{a},\vec{b})\ (i,j=0,1)$ is the joint probability
of measuring results $i$ and $j$ for qubits $a$ and $b$,
respectively.

Note that the minus sign can be moved to any term in Eq. \rqn{4}.
 This is related to the freedom of the interchange of the detectors
for each qubit.
 Indeed, the substitution $\vec{a}\leftrightarrow\vec{a}'$ results in
the interchange of the signs of the second and fourth terms in Eq.
\rqn{4}.
 Similarly, the substitution $\vec{b}\leftrightarrow\vec{b}'$
($\vec{a}\leftrightarrow\vec{a}',\ \vec{b}\leftrightarrow\vec{b}'$)
is equivalent to moving the minus sign to the first (third) term in
Eq. \rqn{4}.

\subsection{The Bell operator}
 \label{IIB}

According to quantum mechanics in the absence of measurement errors
 \be
p_{ij}(\vec{a},\vec{b}) ={\rm Tr}[P_i^a(\vec{a})
P_j^b(\vec{b})\rho].
 \e{13}
Here $\rho$ is the density matrix for the two qubits,
$P_i^a=P_i\otimes I_2$, and $P_i^b=I_2\otimes P_i$, where $I_n$ is
the $n\times n$ identity matrix and $P_i$ are the single-qubit
projection operators,
 \be
P_1(\vec{a})=\frac{1}{2}(I_2+\vec{a}\cdot\vec{\sigma}),\quad
P_0(\vec{a})= \frac{1}{2}(I_2-\vec{a}\cdot\vec{\sigma}),
 \e{29}
$\vec{\sigma}=(\sigma_x,\sigma_y,\sigma_z)$ being the vector of the
Pauli matrices \cite{nie00}.
 Here and below we assume that $|1\rangle$ and $|0\rangle$ are the
eigenvectors of $\sigma_z$ with the eigenvalues 1 and $-1$,
respectively.

Equations \rqn{5} and \rqn{13} yield that $E(\vec{a},\vec{b})={\rm
Tr}(AB\rho)$.
 Here
 \be
A=P_1^a(\vec{a})-P_0^a(\vec{a})= \vec{a}\cdot\vec{\sigma}_a
 \e{3.24}
and similarly $B=\vec{b}\cdot\vec{\sigma}_b$, where
$\vec{\sigma}_a=\vec{\sigma}\otimes I_2$ and
$\vec{\sigma}_b=I_2\otimes\vec{\sigma}$, the eigenvalues of $A$ and
$B$ being $\pm1$.
 Correspondingly, as follows from Eq.\ \rqn{4},
 \be
S={\rm Tr}({\cal B}\rho),
 \e{3.5}
where the Bell operator \cite{bra92} ${\cal B}$ is
 \be
{\cal B}=AB-AB'+A'B+A'B'.
 \e{2.2}
Here $A'=\vec{a}'\cdot\vec{\sigma}_a$ and
$B'=\vec{b}'\cdot\vec{\sigma}_b$.
 Note that measurements along a given direction $\vec{a}$ described
by the projection operators \rqn{29} can be performed by means of a
suitable rotation of the qubit and a subsequent measurement in the
basis $\{|0\rangle,|1\rangle\}$ \cite{kof08,ste06,kat06}.

In this paper we look for experimental conditions which are the most
favorable for an observation of the BI violation.
 Such conditions are reached when $|S|-2$ is positive and maximal.
 We will use the following properties \cite{kof08} of $S$ which
follow from Eqs. \rqn{3.5} and \rqn{2.2}.

 (i) $S$ is invariant under arbitrary local unitary
transformations of the qubits,
 \bes{3.16}
 \be
\rho\rightarrow(U_a\otimes U_b)\rho(U_a^\dagger\otimes U_b^\dagger),
 \e{3.16a}
and the corresponding rotations of the detectors,
 \be
\vec{a}\rightarrow R_a\vec{a},\ \ \vec{a}'\rightarrow R_a\vec{a}', \
\ \vec{b}\rightarrow R_b\vec{b},\ \ \vec{b}'\rightarrow R_b\vec{b}',
 \e{3.16b}
 \ese
where $U_a\ (U_b)$ is a unitary matrix for qubit $a\ (b)$ and $R_a$
($R_b$) is the rotation matrix corresponding to $U_a$ ($U_b$), so
that, e.g., $U_a(\vec{r}_a\cdot\vec{\sigma})U_a^\dagger=
(R_a\vec{r}_a) \cdot\vec{\sigma}$.
 A rotation matrix $R$ is an orthogonal matrix, $R^TR=I_3$, with
$\det(R)=1$.

(ii) $S$ inverts the sign if the vectors $\vec{a}$ and $\vec{a}'$ (or
$\vec{b}$ and $\vec{b}'$) invert the sign (e.g.,
$\vec{a}\rightarrow-\vec{a},\ \vec{a}'\rightarrow-\vec{a}'$).
 Therefore, for a given state the maximal and minimal values of $S$
are equal by the magnitude, yielding equal violations of the both
bounds in the BI \rqn{3}.
 As a result, it is sufficient to discuss only the conditions
for achieving the maximum of $S$.
 Below we denote by $S_+$ and $S_-$ the maximum and minimum,
respectively, of $S$ for a given state and by $S_{\rm max}$ the
maximum of $S$ over all states and detector axes.

\subsection{Maximally entangled states}
 \label{IIC}

 In the ideal case, when there is no decoherence or errors, the
maximal and minimal values, $S_{\rm max}$ and $S_{\rm min}$, which
$S$ can achieve are \cite{cir80}
 \be
S_{\rm max}=2\sqrt{2},\quad S_{\rm min}=-2\sqrt{2}.
 \e{19}
These limits are obtained for any maximally entangled state
\cite{pop92}.
 The BI violations are often considered for the following maximally
entangled states, called also the Bell states \cite{nie00},
 \bea
&&|\Phi_\pm\rangle=(|11\rangle\pm|00\rangle)/\sqrt{2}, \label{3.23}\\
&&|\Psi_\pm\rangle=(|10\rangle\pm|01\rangle)/\sqrt{2}.
 \ea{28}
For each maximally entangled state there are infinitely many optimal
(i.e., producing a maximal BI violation) configurations of the
detector axes $\vec{a}$, $\vec{a}'$, $\vec{b}$, and $\vec{b}'$; all
such configurations were described in Ref. \onlinecite{kof08}.
 In Sec. \ref{III} we describe the configurations maximizing S for an
arbitrary (pure or mixed) state.

It is useful to have a general expression for maximally entangled
states in the ``standard'' basis of the qubit pair.
 It can be shown by performing the Schmidt decomposition of
the general two-qubit state that, with the accuracy to an overall
phase, the most general form of a maximally entangled state is
\cite{note1,vol00}
 \be
|\Psi\rangle=c_1|11\rangle +c_2e^{i\alpha_1}|10\rangle-
c_2e^{i\alpha_2}|01\rangle+c_1e^{i(\alpha_1+\alpha_2)}|00\rangle
 \e{3.26}
Here $\alpha_1$ and $\alpha_2$ are real numbers, whereas
$c_1,c_2\ge0$ and $c_1^2+c_2^2=1/2$.

With the help of local rotations of the qubits around the $z$ axis,
Eq. \rqn{3.26} can be reduced to one of the two wavefunctions
 \be
|\Psi\rangle=c_\Phi|\Phi_\pm\rangle +c_\Psi|\Psi_\mp\rangle,
 \e{3.27}
where either the upper or the lower signs should be used
simultaneously, $c_\Phi$ and $c_\Psi$ are any real numbers satisfying
$c_\Phi^2+c_\Psi^2=1$, and $|\Phi_\pm\rangle$ and $|\Psi_\pm\rangle$
are the Bell states \rqn{3.23} and \rqn{28}.
 Equation \rqn{3.27} is used in Sec. \ref{IV}.

\section{Conditions for the maximal BI violation in a given state}
 \label{III}

 While formulas for the maximum $S_+$ of $S$ are known for both
pure \cite{gis91,pop92a} and mixed \cite{hor95} states, only one
optimal detector configuration (i.e., a configuration for which $S_+$
is realized) was provided in Refs. \onlinecite{hor95,gis91,pop92a}.
 In contrast, Samuelsson et al.\ \cite{sam03} showed that for the
Bell state $|\Phi_+\rangle$ in the presence of dephasing there is a
family of optimal configurations depending on one continuous
parameter.
 In this section we extend the method of Ref.\ \onlinecite{hor95} in
order to obtain all optimal detector configurations for any (pure or
mixed) state.
 We show that the set of optimal detector configurations generally
depends on one continuous and one discrete parameters, though in
special cases the number of continuous parameters can equal two or
three.

\subsection{Real representation of a two-qubit state}
 \label{IIIA}

It is useful to consider the following representation
\cite{hor95,sch95,hio81} of the two-qubit density matrix
 \be
\rho=\left(I_4+\vec{r}_a\cdot\vec{\sigma}_a+
\vec{r}_b\cdot\vec{\sigma}_b+\vec{\sigma}_a{\cal
T}\vec{\sigma}_b\right)/4.
 \e{3.64}
Here $\vec{r}_k$ is the Bloch vector characterizing the reduced
density matrix for the qubit $k$, so that, e.g.,
$\rho_a=\mbox{Tr}_b\rho=(I_2+\vec{r}_a\cdot\vec{\sigma})/2$, $\cal T$
is a matrix with the real elements
 \be
{\cal T}_{mn}=\mbox{Tr}(\rho\sigma_m^a\sigma_n^b)\quad(m,n=x,y,z),
 \e{3.17}
and $\vec{\sigma}_a{\cal T}\vec{\sigma}_b=\sum_{m,n=x}^z{\cal
T}_{mn}\sigma_m^a\sigma_n^b$, where
$\vec{\sigma}_k=(\sigma_x^k,\sigma_y^k,\sigma_z^k)\ (k=a,b)$.

Consider some useful properties of $\cal T$.
 Since the eigenvalues of the Hermitian operators
$\sigma_m^a\sigma_n^b$ equal $\pm1$, Eq. \rqn{3.17} implies that
 \be
|{\cal T}_{mn}|\le1.
 \e{3.65}
 There exists the polar decomposition \cite{nie00}
 \be
{\cal T}={\cal V}\sqrt{\cal U},
 \e{3.30}
where ${\cal V}$ is a $3\times3$ orthogonal matrix, ${\cal V}^T{\cal
V}=I_3$, and ${\cal U}={\cal T}^T{\cal T}$ is a real symmetric matrix
with nonnegative eigenvalues $u_1,\ u_2$, and $u_3$; $u_3$ being the
smallest eigenvalue $(0\le u_3\le u_1,u_2)$.
 ${\cal V}$ is determined uniquely and given by
${\cal V}={\cal TU}^{-1/2}$, only if $u_3\ne0$; this is the most
interesting case, as shown in Sec. \ref{IIIB}.
 The determinant of ${\cal V}$ equals 1 or $-1$ for
$\det({\cal T})>0$ and $\det({\cal T})<0$, respectively; when
$\det({\cal T})=0$ (which means that $u_3=0$), ${\cal V}$ can be
chosen such that $\det({\cal V})=1$.

Under a local unitary transformation $\rho\rightarrow(U_a\otimes
U_b)\rho(U_a^\dagger\otimes U_b^\dagger)$ Eq. \rqn{3.64} changes so
that $\vec{r}_k\rightarrow R_k\vec{r}_k$ and \cite{sch95}
 \be
{\cal T}\rightarrow R_a{\cal T}R_b^T,
 \e{3.47}
where $R_{a,b}$ are defined after Eq. \rqn{3.16b}.
 As follows from Eqs. \rqn{3.30} and \rqn{3.47}, with suitable rotations
of the qubits, $R_a'$ and $R_b'$, the matrix ${\cal T}$ can be
reduced to one of the two diagonal forms, ${\cal T}'=R_a'{\cal
T}(R_b')^T=\pm\sqrt{\cal U'}$.
 Here $R_b'$ is such that ${\cal U'}=R_b'{\cal U}(R_b')^T$ is
diagonal and $R_a'=\pm R_b'{\cal V}^T$.
 The plus and minus signs in the above formulas are obtained for
$\det({\cal T})\ge0$ and $\det({\cal T})<0$, respectively [on
choosing $\det({\cal V})=1$ when $\det({\cal T})=0$].
 As a consequence, in view of Eq. \rqn{3.65}, we obtain
 \be
0\le u_3\le u_1,u_2\le1.
 \e{3.66}

Note that an arbitrary two-qubit state reduces to a simpler form by a
local unitary transformation which diagonalizes ${\cal T}$.
 Inserting a general diagonal ${\cal T}$ into Eq. \rqn{3.64}, we obtain
that all states with a diagonal ${\cal T}$ have the form
 \be
\rho=\left(\begin{array}{cccc}
 \rho_{11}&\rho_{12}&\rho_{13}&\rho_{14}\\
\rho_{21}&\rho_{22}&\rho_{23}&\rho_{13}\\
\rho_{31}&\rho_{23}&\rho_{33}&\rho_{12}\\
\rho_{14}&\rho_{31}&\rho_{21}&\rho_{44}
\end{array}\right),
 \e{3.69}
where $\rho_{ij}$ subscript values $i,j=1,2,3,4$ correspond to the
basis $\{|11\rangle,|10\rangle,|01\rangle,|00\rangle\}$.
 In the state \rqn{3.69} $\rho_{12}=\rho_{34}$ and
$\rho_{13}=\rho_{24}$, whereas $\rho_{14}$ and $\rho_{23}$ are real
\cite{note3}.

 In view of Eq. \rqn{3.17}, for the state \rqn{3.69} we obtain
 \be
{\cal T}=2\,\mbox{diag}(\rho_{23}+\rho_{14},\rho_{23}-\rho_{14},
1/2-\rho_{22}-\rho_{33}).
 \e{3.71}
Note that this expression is independent of $\rho_{12}$ and
$\rho_{13}$.

\subsection{Maximal BI violation}
 \label{IIIB}

Let us review the derivation \cite{hor95} of the maximum $S_+$ of $S$
for a given state.
 Inserting Eq. \rqn{2.2} into \rqn{3.5} and taking into account Eqs.
\rqn{3.24} and \rqn{3.17}, we obtain that
 \be
S=\vec{a}{\cal T}\vec{b}-\vec{a}{\cal T}\vec{b}' +\vec{a}'{\cal
T}\vec{b}+\vec{a}'{\cal T}\vec{b}'.
 \e{3.6}
The vectors $\vec{b}$ and $\vec{b}'$ can be always written in the
form \cite{pop92a}
 \bea
&&\vec{b}=\vec{c}\,_1'\cos(\zeta_b/2)+\vec{c}\,_2'\sin(\zeta_b/2),
\nonumber\\
&&\vec{b}'=\vec{c}\,_1'\cos(\zeta_b/2)-\vec{c}\,_2'\sin(\zeta_b/2),
 \ea{3.9}
where $\vec{c}\,_1'$ and $\vec{c}\,_2'$ are orthogonal unit vectors
and $\zeta_b$ is the angle between $\vec{b}$ and $\vec{b}'\
(0<\zeta_b<\pi)$.
 Inserting Eq. \rqn{3.9} into \rqn{3.6} yields
$S=2[\vec{a}{\cal T}\vec{c}\,_{2}'\sin(\zeta_b/2) +\vec{a}'{\cal
T}\vec{c}\,_1'\cos(\zeta_b/2)]$.
 To maximize this expression, one should require $\vec{a}$ and
$\vec{a}'$ to be parallel to ${\cal T}\vec{c}\,_2'$ and ${\cal
T}\vec{c}\,_1'$, respectively \cite{note}, yielding
 \be
\vec{a}={\cal T}\vec{c}\,_2'/|{\cal T}\vec{c}\,_2'|,\quad
\vec{a}'={\cal T}\vec{c}\,_1'/|{\cal T}\vec{c}\,_1'|.
 \e{3.10}
Then maximizing $S$ over $\zeta_b$ results in
 \be
\zeta_b=2\arctan(|{\cal T}\vec{c}\,_2'|/|{\cal T}\vec{c}\,_1'|),
 \e{5.23}
where $|\vec{v}|$ denotes the length of a vector $\vec{v}$, and
$S=2\sqrt{|{\cal T}\vec{c}\,_1'|^2+ |{\cal
T}\vec{c}\,_2'|^2}=2\sqrt{\vec{c}\,_1'{\cal
U}\vec{c}\,_1'+\vec{c}\,_2'{\cal U}\vec{c}\,_2'}$.
 The maximum of $S$ is obtained when (see Sec. \ref{IIIC})
 \be
\vec{c}\,_1'{\cal U}\vec{c}\,_1'+\vec{c}\,_2'{\cal
U}\vec{c}\,_2'=u_1+u_2,
 \e{3.3}
yielding \cite{hor95}
 \be
S_+=2\sqrt{u_1+u_2}.
 \e{3.11}

Hence, the BI violation, $S_+>2$, occurs when $u_1+u_2>1$.
 Equations \rqn{3.66} and \rqn{3.11} imply the limits on $S_+$,
 \be
0\le S_+\le2\sqrt{2}.
 \e{3.67}
Here the lower limit, $S_+=0$, is obtained for the states with ${\cal
T}=0$, which are, in view of Eq. \rqn{3.71}, the states locally
equivalent to
$\rho=\mbox{diag}(\rho_{11},\rho_{22},1/2-\rho_{22},1/2-\rho_{11})$.
 [In this case the elements $\rho_{12}$ and $\rho_{13}$ in Eq.
\rqn{3.69} can be set to zero by local rotations of the vectors
$\vec{r}_a$ and $\vec{r}_b$ in Eq. \rqn{3.64} to the $z$ axis.]
 Such states can be shown to be mixtures of product states with one of
the qubits in the maximally mixed state $I_2/2$.

 As an example, let us apply Eq. \rqn{3.11} to the state \rqn{3.69}.
 In view of Eq. \rqn{3.71}, ${\cal U}={\cal T}^2$, yielding
 \bea
&S_+=&2[\max\{8\rho_{23}^2+8\rho_{14}^2,4(|\rho_{23}|-|\rho_{14}|)^2
\nonumber\\
&&+(1-2\rho_{22}-2\rho_{33})^2\}]^{1/2}.
 \ea{3.70}

 The states with $\det({\cal T})\ge0$ do not violate the BI.
 To show this, it is sufficient to consider a diagonal
${\cal T}=\sqrt{\cal U}$, since, as mentioned in Sec. \ref{IIIA},
such ${\cal T}$ can be obtained for any state with $\det({\cal
T})\ge0$ by means of local unitary transformations, which do not
change $S$.
 A diagonal ${\cal T}=\sqrt{\cal U}$ is given by Eq. \rqn{3.71} with
nonnegative matrix elements, which implies that
$r\equiv\rho_{22}+\rho_{33}\le1/2$ and $\rho_{23}\ge|\rho_{14}|$.
 As follows from Eq. \rqn{3.71}, $\mbox{Tr}\,{\cal U}=
\mbox{Tr}\,{\cal T}^2=4\rho_{14}^2 +4\rho_{23}^2+(1-2r)^2$.
 For a given $r$, $\mbox{Tr}\,{\cal U}$ is maximal if $\rho_{23}$ and
$|\rho_{14}|$ are maximal under the above constraints, i.e., if
$|\rho_{14}|=\rho_{23}=r/2$, which yields $\mbox{Tr}\,{\cal
U}\le6r^2-4r+1\ (0\le r\le1/2)$.
 This expression achieves the maximum $\mbox{Tr}\,{\cal U}=1$ for
$r=0$.
 Thus, in the case $\det({\cal T})\ge0$ we have
$\mbox{Tr}\,{\cal U}\le1$, the value $\mbox{Tr}\,{\cal U}=1$ being
obtained for the states which, with the accuracy to local unitary
transformations, have the form
$\rho=\mbox{diag}(\rho_{11},0,0,\rho_{44})$ and hence have ${\cal
T}={\cal U}=\mbox{diag}(0,0,1)$ [see Eq. \rqn{3.71}].
 In view of Eq. \rqn{3.11}, these states yield $S_+=2$; they include,
in particular, pure nonentangled states.

However, the BI violation implies $\mbox{Tr}\,{\cal U}\ge u_1+u_2>1$.
 Hence, a necessary condition for the BI violation is
$\det({\cal T})<0$.
 As a consequence, in view of Eq. \rqn{3.30}, for the states violating
the BI all $u_i$ do not vanish.

\subsection{Optimal detector configurations}
 \label{IIIC}

Consider detector configurations providing the maximal $S$
\rqn{3.11}.
 Equation \rqn{3.3} obviously holds for $\vec{c}\,_1'=\vec{c}_1$ and
$\vec{c}\,_2'=\vec{c}_2$, where $\vec{c}_i\ (i=1,2,3)$ are the unit
orthogonal eigenvectors of ${\cal U}$ corresponding to the
eigenvalues $u_i$.
 This is the choice of $\vec{c}\,_1'$ and $\vec{c}\,_2'$ which was
made in Ref. \onlinecite{hor95} (similar choices were made also in
Refs. \onlinecite{gis91} and \onlinecite{pop92a}).
 In this case Eq. \rqn{5.23} yields $\zeta_b=\zeta_0$, where
 \be
\zeta_0=2\arctan\sqrt{u_2/u_1},
 \e{3.4}
since $|{\cal T}\vec{c}_i|^2=\vec{c}_i{\cal
U}\vec{c}_i=u_i\vec{c}_i\cdot\vec{c}_i=u_i$.
 Hence, Eq. \rqn{3.9} becomes
 \bea
&&\vec{b}=\vec{c}_1\cos(\zeta_0/2)+\vec{c}_2\sin(\zeta_0/2),
\nonumber\\
&&\vec{b}'=\vec{c}_1\cos(\zeta_0/2)-\vec{c}_2\sin(\zeta_0/2),
 \ea{3.50}
and, in view of Eqs. \rqn{3.10} and \rqn{3.30},
 \be
\vec{a}=\vec{e}_2,\quad\vec{a}'=\vec{e}_1,
 \e{3.40}
where
 \be
\vec{e}_i={\cal T}\vec{c}_i/\sqrt{u_i}={\cal V}\vec{c}_i\quad
(i=1,2).
 \e{3.74}
The vectors $\vec{e}_{1,2}$ are orthonormal, since ${\cal V}$ is an
orthogonal matrix.

The optimal detector configuration given by Eqs. \rqn{3.50} and
\rqn{3.40} is not unique.
 To obtain all possible detector configurations providing Eq.
\rqn{3.11}, we consider the derivation of the maximum \rqn{3.3}, as
follows.
 If $\vec{c}\,_3'$ is a unit vector orthogonal to $\vec{c}\,_1'$ and
$\vec{c}\,_2'$, then $\vec{c}\,_i'=\sum_{j=1}^3{\cal
W}_{ij}\vec{c}_j\ (i=1,2,3)$, where ${\cal W}$ is an orthogonal
$3\times3$ matrix, ${\cal W}^T{\cal W}=I_3$.
 We have $\sum_{i=1}^3\vec{c}\,_i'{\cal U}\vec{c}\,_i'
=\sum_{i,j,k=1}^3{\cal W}_{ij}{\cal W}_{ik}\vec{c}\,_j{\cal
U}\vec{c}\,_k=\sum_{j,k=1}^3({\cal W}^T{\cal W})_{jk}\vec{c}\,_j{\cal
U}\vec{c}\,_k =\sum_{j=1}^3\vec{c}\,_j{\cal
U}\vec{c}\,_j=u_1+u_2+u_3$.
 Moreover,
 \be
\vec{c}\,_3'{\cal U}\vec{c}\,_3'={\cal W}_{31}^2u_1+{\cal
W}_{32}^2u_2+{\cal W}_{33}^2u_3\ge u_3,
 \e{3.33}
since ${\cal W}_{31}^2+{\cal W}_{32}^2+{\cal W}_{33}^2=1$ and $u_3\le
u_1+u_2$.
 As a result, $\vec{c}\,_1'{\cal U}\vec{c}\,_1'+
\vec{c}\,_2'{\cal U}\vec{c}\,_2' =u_1+u_2+u_3-\vec{c}\,_3'{\cal
U}\vec{c}\,_3'\le u_1+u_2$.
 Hence, the maximum \rqn{3.3} is achieved when the expression \rqn{3.33}
is minimal, which occurs for ${\cal W}_{31}={\cal W}_{32}=0$ and
${\cal W}_{33}=\pm1$, i.e., for $\vec{c}\,_3'=\pm\vec{c}_3$.
 In this case $\vec{c}\,_1'$ and $\vec{c}\,_2'$ are an arbitrary
pair of orthonormal vectors in the plane defined by $\vec{c}_1$ and
$\vec{c}_2$.
 All such $\vec{c}\,_1'$ and $\vec{c}\,_2'$ are given by
 \be
\vec{c}\,_1'=\vec{c}_1\cos\eta\pm\vec{c}_2\sin\eta,\ \
\vec{c}\,_2'=-\vec{c}_1\sin\eta\pm\vec{c}_2\cos\eta.
 \e{5.21}
Here $\eta$ is the arbitrary angle of rotation of $\vec{c}_1$ and
$\vec{c}_2$, and the two signs before $\vec{c}_2$ correspond to a
reflection with respect to the $\vec{c}_1$ axis.
 The above derivation implies that no pair of unit orthogonal vectors
$\vec{c}\,_1'$ and $\vec{c}\,_2'$ other than those in Eq. \rqn{5.21}
can satisfy the condition \rqn{3.3}, unless $u_3$ equals $u_1$ or
$u_2$ (the latter cases are discussed in Sec. \ref{IIIF}).

Inserting Eq. \rqn{5.21} into Eqs. \rqn{3.9} and \rqn{3.10} and
taking into account Eq. \rqn{3.74}, we obtain
 \bea
&&\vec{a}=(-\vec{e}_1\sqrt{u_1}\sin\eta\pm
\vec{e}_2\sqrt{u_2}\cos\eta)
/|{\cal T}\vec{c}\,_2'|,\nonumber\\
&& \vec{a}'=(\vec{e}_1\sqrt{u_1}\cos\eta\pm
\vec{e}_2\sqrt{u_2}\sin\eta)/|{\cal T}\vec{c}\,_1'|,\nonumber\\
&&\vec{b}=\vec{c}_1\cos[\eta+\zeta_b(\eta)/2]\pm\vec{c}_2
\sin[\eta+\zeta_b(\eta)/2],
\nonumber\\
&&\vec{b}'=\vec{c}_1\cos[\eta-\zeta_b(\eta)/2]\pm\vec{c}_2
\sin[\eta-\zeta_b(\eta)/2],
 \ea{5.22}
In Eq. \rqn{5.22} the upper (or lower) signs should be used
simultaneously.
 The quantities $|{\cal T}\vec{c}\,_1'|$ and
$|{\cal T}\vec{c}\,_2'|$ are given by
 \be
|{\cal T}\vec{c}\,_{1(2)}'|=\sqrt{[u_1+u_2\pm(u_1-u_2)\cos2\eta]/2},
 \e{3.2}
where the upper (lower) sign corresponds to $|{\cal T}\vec{c}\,_1'|$
($|{\cal T}\vec{c}\,_2'|$).
 Inserting Eq. \rqn{3.2} into Eq. \rqn{5.23} and performing trigonometric
calculations yields
 \be
\zeta_b(\eta)=\arccos\left(\frac{u_1-u_2}{u_1+u_2}\cos2\eta\right).
 \e{3.7}

Equation \rqn{3.7} can be compared with the angle $\zeta_a$ between
$\vec{a}$ and $\vec{a}'$, satisfying
$\cos\zeta_a=\vec{a}\cdot\vec{a}'$.
 As follows from Eqs. \rqn{5.22} and \rqn{3.2} and some calculations,
 \be
\zeta_a(\eta)={\rm
arccot}\left(\frac{u_2-u_1}{2\sqrt{u_1u_2}}\sin2\eta\right)
 \e{3.8}
(we assume that $0<\zeta_a,\zeta_b<\pi$).
 Equations \rqn{3.7} and \rqn{3.8} imply that the angles $\zeta_a$ and
$\zeta_b$ vary with $\eta$ between the values $\zeta_0$ \rqn{3.4} and
$\pi-\zeta_0$ with the period $\pi$.
 In particular, for $\eta=0,\pm\pi/2,\pm\pi$
($\eta=\pm\pi/4,\pm3\pi/4$) one gets $\zeta_a=\pi/2$
($\zeta_b=\pi/2$), whereas $\zeta_b$ ($\zeta_a$) acquires a maximal
or minimal value.
 Moreover, Eqs. \rqn{5.22}-\rqn{3.7} imply the relations
 \be
\vec{a}'(\eta\pm\pi/2)=\pm\vec{a}(\eta),\quad
\vec{b}'(\eta\pm\pi/2)=\pm\vec{b}(\eta),
 \e{3.21}
whereas $\vec{a},\vec{a}',\vec{b}$, and $\vec{b}'$ change the sign
for $\eta\rightarrow\eta+\pi$.

Note that the optimal detector orientations \rqn{5.22} depend on
$u_1$ and $u_2$ only through the ratio $u_2/u_1$.
 As a result, two different states with the same ${\cal V},\
\vec{c}_1,\ \vec{c}_2$, and $u_2/u_1$ have the same or, at least,
overlapping sets of optimal detector configurations.
 (The overlap may be incomplete only when, at least, for one of the
states $u_1=u_3$ or $u_2=u_3$, see Sec. \ref{IIIF}.)
 In particular, for the states $\rho$ and $\rho'$ with the matrices
${\cal T}$ and ${\cal T}'$, respectively, satisfying ${\cal
T}'=f{\cal T}\ (f>0)$, the respective quantities $S_+$ and $S_+'$
obey $S_+'=fS_+$, whereas the optimal configurations are the same for
the both states.
 An example of such states $\rho$ and $\rho'$ are, respectively, the
input and output states of the depolarizing channel \cite{nie00},
 \be
\rho'=f\rho+(1-f)(I_4/4)\quad(0<f\le1).
 \e{3.73}

\subsection{Polar coordinates}
\label{IID}

 According to Eq. \rqn{5.22}, the optimal detector directions for
qubits $a$ and $b$ are confined to the planes $(\vec{e}_1,\vec{e}_2)$
and $(\vec{c}_1,\vec{c}_2)$, respectively.
 It is convenient to specify these directions by means of
polar angles.

 To this end, we introduce the polar coordinate $\nu$ in the
$(\vec{c}_1,\vec{c}_2)$ plane, which is counted from $\vec{c}_1$ in
the direction where $\nu=\pi/2$ corresponds to $\vec{c}_2$, and the
polar coordinate $\delta$ in the $(\vec{e}_1,\vec{e}_2)$ plane
[differing generally from the $(\vec{c}_1,\vec{c}_2)$ plane], which
is counted from $\vec{e}_1$ in the direction where $\delta=\pi/2$ is
the polar coordinate of $\vec{e}_2$.
 Then Eq. \rqn{5.22} can be recast in the form
 \bea
&&\vec{a}=\vec{e}_1\cos[\delta_a(\eta)]\pm
\vec{e}_2\sin[\delta_a(\eta)],\nonumber\\
&&\vec{a}'=\vec{e}_1\cos[\delta_a'(\eta)]\pm
\vec{e}_2\sin[\delta_a'(\eta)],\nonumber\\
&&\vec{b}=\vec{c}_1\cos[\nu_b(\eta)]\pm\vec{c}_2\sin[\nu_b(\eta)],
\nonumber\\
&&\vec{b}'=\vec{c}_1\cos[\nu_b'(\eta)]\pm\vec{c}_2
\sin[\nu_b'(\eta)].
 \ea{3.22}
Here $\delta_a,\ \delta_a',\ \nu_b$, and $\nu_b'$ ($-\delta_a,\
-\delta_a',\ -\nu_b$, and $-\nu_b'$) are the polar coordinates of the
vectors $\vec{a},\ \vec{a}',\ \vec{b}$, and $\vec{b}'$, respectively,
when the upper (lower) sign in Eq. \rqn{3.22} is realized.
 The configurations corresponding to the two different choices of the
sign in Eq. \rqn{5.22} or \rqn{3.22} transform to each other by the
reflection of the detector axes for qubits $a$ and $b$ with respect
to the axes $\vec{e}_1$ and $\vec{c}_1$, respectively.

As follows from Eq. \rqn{3.21}, the functions $\delta_a(\eta),\
\delta_a'(\eta),\ \nu_b(\eta)$, and $\nu_b'(\eta)$ satisfy the
relations
 \be
\delta_a(\eta)=\delta_a'(\eta+\pi/2),\quad
\nu_b(\eta)=\nu_b'(\eta+\pi/2).
 \e{3.14}
Expressions for these functions are obtained by comparing Eqs.
\rqn{5.22} and \rqn{3.22}.
 In particular, we obtain
 \be
\nu_b(\eta)=\eta+\zeta_b(\eta)/2,
\quad\nu_b'(\eta)=\eta-\zeta_b(\eta)/2.
 \e{3.12}
This implies that $\eta=(\nu_b+\nu_b')/2$, yielding a geometric
interpretation of $\eta$: $\eta$ equals the polar coordinate of the
bisector of the angle between $\vec{b}$ and $\vec{b}'$ when the upper
sign in Eq. \rqn{5.22} or \rqn{3.22} is realized.

Let us now obtain $\delta_a'(\eta)$; then $\delta_a(\eta)$ will
follow by the first Eq. \rqn{3.14}.
 The second Eq. \rqn{5.22} implies that
$\cos\delta_a'=\vec{a}'\cdot \vec{e}_1= \sqrt{u_1}\cos\eta/|{\cal
T}\vec{c}\,_1'|$, whereas the sign of $\sin\delta_a'$ coincides with
the sign of $\sin\eta$.
 With the account of Eq. \rqn{3.2}, this yields (for
$-\pi/2\le\eta\le\pi/2$)
 \be
\delta_a'(\eta)=\arctan\left(\sqrt{u_2/u_1}\tan\eta\right),\ \
\delta_a(\eta)=\delta_a'(\eta+\pi/2).
 \e{3.15}
It is convenient to consider the polar angles as continuous functions
of $\eta$ without restricting them to any interval.
 In particular, $\delta_a'(\eta)$ in Eq. \rqn{3.15} can be extended
continuously beyond the interval $-\pi/2\le\eta\le\pi/2$ with the
help of the equality $\delta_a'(\eta+k\pi)=\delta_a'(\eta)+k\pi\
(k=\pm1,\pm2,\dots)$.

\begin{figure}[htb]
\includegraphics[width=8.5cm]{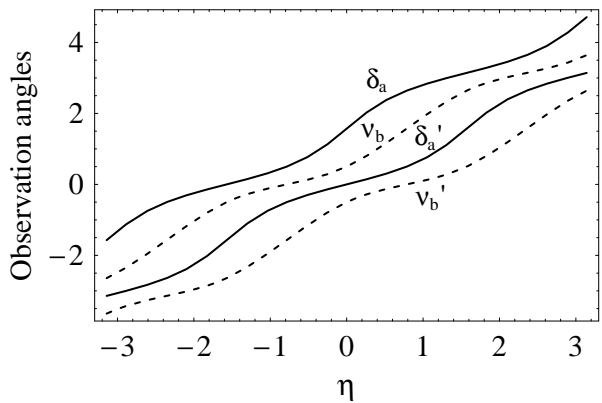}
\caption{Polar angles of the detector axes versus $\eta$ for
$u_2/u_1=0.3$ ($\zeta_0\approx1.00$), as given by Eqs.
\protect\rqn{3.12} and \protect\rqn{3.15}.}
 \label{f1}\end{figure}

Figure \ref{f1} shows the dependence on $\eta$ of the polar
coordinates of the observation axes when the upper sign is chosen in
Eq. \rqn{3.22} and $u_2/u_1=0.3$ ($\zeta_0\approx1.00$).
 Note that in Fig. \ref{f1} the observation angles are not restricted
to an interval of the length $2\pi$ to avoid discontinuities in the
plots.
 Our calculations show that the set of the polar coordinates of the
detectors taken as continuous functions of $\nu$ is always ordered as
follows (cf. Fig. \ref{f1}),
 \be
\delta_a>\nu_b>\delta_a'>\nu_b'.
 \e{3.45}

\subsection{Symmetry with respect to the exchange of the qubits}

In the above derivation (Sec. \ref{IIIB}) the detectors for qubit $b$
are treated differently than those for qubit $a$ [cf. Eqs. \rqn{3.9}
and \rqn{3.10}], resulting in apparently different solutions for the
two qubits [cf. Eqs. \rqn{3.7} and \rqn{3.8}].
 It is of interest to check whether these solutions are symmetric
with respect to the qubit swap.
 We cast Eq. \rqn{3.6} as
 \be
S=\vec{b}{\cal T}^T\vec{a}-\vec{b}{\cal T}^T\vec{a}' +\vec{b}'{\cal
T}^T\vec{a}+\vec{b}'{\cal T}^T\vec{a}'.
 \e{3.29}
As follows from Eq. \rqn{3.30}, $\vec{b}{\cal T}^T\vec{a}=
\vec{b}{\cal V}^T{\cal T}{\cal V}^T\vec{a} =({\cal V}\vec{b}){\cal
T}({\cal V}^T\vec{a})$.
 Thus, Eq. \rqn{3.29} can be recast as
 \bea
&S=&({\cal V}\vec{b}){\cal T}({\cal V}^T\vec{a})-
({\cal V}\vec{b}'){\cal T}({\cal V}^T\vec{a})\nonumber\\
&&+({\cal V}\vec{b}){\cal T}({\cal V}^T\vec{a}')+({\cal
V}\vec{b}'){\cal T}({\cal V}^T\vec{a}').
 \ea{3.31}
Comparing Eqs. \rqn{3.6} and \rqn{3.31}, we obtain that $S$ is
invariant under the simultaneous substitutions
 \be
{\cal V}\vec{b}'\rightarrow\vec{a},\ \ {\cal
V}\vec{b}\rightarrow\vec{a}',\ \ {\cal
V}^T\vec{a}'\rightarrow\vec{b},\ \ {\cal
V}^T\vec{a}\rightarrow\vec{b}'.
 \e{3.32}

In view of Eq. \rqn{3.32}, we obtain that the qubit-swap operation is
equivalent to the replacement
 \be
\delta_a\rightarrow\nu_b',\ \ \delta_a'\rightarrow\nu_b,\ \
\nu_b\rightarrow\delta_a',\ \ \nu_b'\rightarrow\delta_a
 \e{3.44}
in Eq. \rqn{3.22}.
 As shown in Sec. \ref{IIC}, Eq. \rqn{5.22} or \rqn{3.22} describes all
optimal detector configurations (at least, for $u_3<u_1,u_2$).
 Therefore, optimal configurations transformed by the replacement
\rqn{3.44} should belong to the set of detector configurations
described by Eq. \rqn{5.22} or \rqn{3.22}.
 This is confirmed by our numerical calculations which show that
$\delta_a,\ \delta_a'$, and $\nu_b$ as functions of $\nu_b'$
coincide, respectively, with $-\nu_b',\ -\nu_b$, and $-\delta_a'$ as
functions of $-\delta_a$.
 In other words, Eqs. \rqn{3.12} and \rqn{3.15} imply that if
$(\delta_a,\delta_a',\nu_b,\nu_b')$ are optimal polar angles of the
vectors $\vec{a},\ \vec{a}',\ \vec{b}$, and $\vec{b}'$, respectively,
then $(-\nu_b',-\nu_b,-\delta_a',-\delta_a)$ [and hence
$(\nu_b',\nu_b,\delta_a',\delta_a)$] are also optimal polar angles
for the respective vectors.
 Thus, the set of the detector configurations \rqn{5.22} or \rqn{3.22} is
invariant with respect to the qubit-swap symmetry relation \rqn{3.32}
[or \rqn{3.44}].

As follows from the above, the set of all optimal configurations
maximizing $S$ for a given state generally depends on one continuous
parameter ($\eta$) and one discrete parameter [which corresponds to
the two possible signs in Eq. \rqn{5.22} or \rqn{3.22}].
 However, when $u_3$ is equal to $u_1$ and/or $u_2$ (a degenerate
case), the set of optimal configurations is characterized by two or
three continuous parameters, as discussed in Sec \ref{IIIF}.

\subsection{Degenerate cases}
\label{IIIF}

Consider the optimal detector configurations in degenerate cases.

\paragraph{The singlet state.}
We begin with the singlet state $\rho=|\Psi_-\rangle\langle\Psi_-|$
[Eq. \rqn{28}], which is often considered in connection with the BI.
 In this case Eq. \rqn{3.17} yields an especially simple result,
${\cal T}=-I_3$ and ${\cal U}={\cal T}^2=I_3$, implying
$u_1=u_2=u_3=1$.
 Hence, in view of Eq. \rqn{3.11}, $S_+=2\sqrt{2}$ [cf. Eq. \rqn{19}].
 Now any two perpendicular unit vectors can be chosen as $\vec{c}_1$
and $\vec{c}_2$.
 Since ${\cal V}={\cal T}=-I_3$ [see Eq. \rqn{3.30}] and
$\zeta_0=\pi/2$ [see Eq. \rqn{3.4}], in view of Eqs. \rqn{3.40} and
\rqn{3.50}, we obtain $\vec{a}=-\vec{c}_2$, $\vec{a}'=-\vec{c}_1$,
and
 \be
\vec{b}=-(\vec{a}+\vec{a}')/\sqrt{2},\quad
\vec{b}'=(\vec{a}-\vec{a}')/\sqrt{2}.
 \e{g2}
Equation \rqn{g2} with arbitrary mutually perpendicular $\vec{a}$ and
$\vec{a}'$ is known to provide all optimal configurations for the
singlet state \cite{kof08}.
 They are characterized by three continuous
parameters, since all optimal configurations can be obtained by
arbitrary rotations of any given optimal configuration \cite{kof08}.

\paragraph{Maximally entangled states.}
 Any maximally entangled state is obtained from the singlet by
local rotations of the qubits, so that, in view of Eq. \rqn{3.47},
for such a state ${\cal T}=-R$, where $R=R_aR_b^T$ corresponds to the
rotation of qubit $a$ relative to qubit $b$.
 In this case $\vec{r}_a=\vec{r}_b=0$, and hence, in view of Eq.
\rqn{3.64}, $\rho=(I_4-\vec{\sigma}_a R\vec{\sigma}_b)/4$, i.e., a
maximally entangled state is determined \cite{vol00} by $R$.

 Now ${\cal U}=R^TR=I_3$, as for the singlet.
 Proceeding as in the case of the singlet, we obtain
$\vec{a}=-R\vec{c}_2$, $\vec{a}'=-R\vec{c}_1$.
 As a result, all optimal configurations providing $S_+=2\sqrt{2}$
are given by Eq. \rqn{g2} with $\vec{a}\rightarrow R^T\vec{a}$ and
$\vec{a}'\rightarrow R^T\vec{a}'$, so that
 \be
\vec{b}=-(R^T\vec{a}+R^T\vec{a}')/\sqrt{2},\quad
\vec{b}'=(R^T\vec{a}-R^T\vec{a}')/\sqrt{2}.
 \e{3.48}
Since the rotation matrix $R$ is determined by three parameters
(e.g., the Euler angles), all optimal configurations for all
maximally entangled states are characterized by six continuous
parameters \cite{kof08}.

 Note in passing that all configurations producing $S_-=-2\sqrt{2}$
follow from Eq. \rqn{3.48} with $\vec{a}\rightarrow-\vec{a}$ and
$\vec{a}'\rightarrow-\vec{a}'$ (see Sec. \ref{IIB}), yielding
 \be
\vec{b}=(R^T\vec{a}+R^T\vec{a}')/\sqrt{2},\quad
\vec{b}'=(R^T\vec{a}'-R^T\vec{a})/\sqrt{2}.
 \e{3.52}

\paragraph{Completely degenerate case ($u_1=u_2=u_3$).}
 Let ${\cal U}=uI_3$, yielding $u_1=u_2=u_3=u$.
 In this case Eq. \rqn{3.11} yields $S_+=2\sqrt{2u}$, so that the BI is
violated for
 \be
u>1/2.
 \e{3.49}
Now ${\cal T}=\sqrt{u}{\cal V}$, where ${\cal V}$ is an orthogonal
matrix [see Eq. \rqn{3.30}].
 When the BI is violated [and hence Eq. \rqn{3.49} holds],
$\det({\cal T})<0$ and hence $R=-{\cal V}$ is a rotation matrix.
 In this case the optimal configurations are the same as for the
maximally entangled state, Eq. \rqn{3.48} (cf. the last paragraph of
Sec. \ref{IIIC}).
 As a result, in the completely degenerate case the set of optimal
configurations is characterized by three continuous parameters.

An example of a state corresponding to the completely degenerate case
is the Werner state \cite{wer89}
 \be
\rho=\sqrt{u}|\Psi_-\rangle\langle\Psi_-|+(1-\sqrt{u})(I_4/4),
 \e{3.51}
which is a special case of the state \rqn{3.73}.
 The BI violation condition \rqn{3.49} for this state was obtained in
Ref. \onlinecite{hor95}.
 For the state \rqn{3.51} ${\cal T}=-\sqrt{u}I_3$, yielding $R=I_3$,
and we obtain the same optimal configurations \rqn{g2} as for the
singlet.

\paragraph{Case $u_1=u_3$ or $u_2=u_3$.}
 In this case we assume for definiteness that $u_2=u_3$.
 Then $\vec{c}_1$ is defined uniquely, whereas $\vec{c}_2$ can be any
unit radius-vector in the plane perpendicular to $\vec{c}_1$.
 Now the set of optimal configurations \rqn{5.22} or \rqn{3.22} is
characterized by two continuous parameters, $\eta$ and an angle
specifying the direction of $\vec{c}_2$ in the plane
$(\vec{c}_2,\vec{c}_3)$ with respect to some reference axis.
 [In this case the discrete parameter is superseded by the new
continuous parameter; indeed, now one of the two possible signs in
Eqs. \rqn{5.22} and \rqn{3.22} can be omitted, since it is recovered
for $\vec{c}_2\rightarrow-\vec{c}_2$.]

\paragraph{Case $u_1=u_2$.}
 In the case $u_1=u_2=u$ Eq. \rqn{3.11} yields $S_+=2\sqrt{2u}$, as
in the completely degenerate case.
 Equations \rqn{3.2}-\rqn{3.8} yield $|{\cal
T}\vec{c}\,_1'|= |{\cal T}\vec{c}\,_2'|=\sqrt{u}$ and
 \be
\zeta_a=\zeta_b=\pi/2.
 \e{3.28}
Now any two perpendicular unit vectors can be chosen as $\vec{c}_1$
and $\vec{c}_2$ in the plane spanned by the eigenvectors of ${\cal
U}$ corresponding to $u_1$ and $u_2$.
 Hence, as follows from Eqs. \rqn{3.50} and \rqn{3.40}, the optimal
configurations are given by Eq. \rqn{3.48}, where $R=-{\cal V}$ and
$\vec{a}$ and $\vec{a}'$ are arbitrary perpendicular unit vectors in
the plane $(\vec{e}_1,\vec{e}_2)$.
 In other words, now in the case of interest $\det({\cal V})=-1$ the
optimal configurations are a subset of the set of the optimal
configurations for the maximally entangled state characterized by the
rotation matrix $R=-{\cal V}$.
 The optimal configurations are described by the following polar
angles [see Eqs. \rqn{3.12} and \rqn{3.15}],
 \be
\delta_a=\eta+\pi/2,\ \delta_a'=\eta,\ \nu_b=\eta+\pi/4,\
\nu_b'=\eta-\pi/4.
 \e{3.13}
Now the set of optimal configurations is characterized by one
continuous parameter and one discrete parameter as in the
nondegenerate case.

\subsection{Special cases}
\label{IIIG}

Here we consider the states with ${\cal T}$ assuming one of the two
simple forms,
 \bea
&&{\cal T}=\mbox{diag}(\tau_x,\tau_x,-\tau_z),\label{3.54}\\
&&{\cal T}=\mbox{diag}(\tau_x,-\tau_x,\tau_z).
 \ea{3.56}
In view of Eq. \rqn{3.71}, the most general two-qubit states with
${\cal T}$ given by Eq. \rqn{3.54} or \rqn{3.56} are described by Eq.
\rqn{3.69} with $\rho_{14}=0$ or $\rho_{23}=0$, respectively, so that
in Eq. \rqn{3.54} [\rqn{3.56}] $\tau_x=2\rho_{23}$ and
$\tau_z=2\rho_{22}+2\rho_{33}-1$ ($\tau_x=2\rho_{14}$ and
$\tau_z=1-2\rho_{22}-2\rho_{33}$).
 We assume below the validity of a necessary condition for the
BI violation, $\det({\cal T})<0$ (Sec. \ref{IIIB}), which implies
$\tau_x\ne0$ and $\tau_z>0$.
 The latter inequality is equivalent to the conditions
$\rho_{22}+\rho_{33}>1/2$ and $\rho_{22}+\rho_{33}<1/2$ for the cases
\rqn{3.54} and \rqn{3.56}, respectively.
 Without a loss of generality, we focus on the states with
$\tau_x>0$ in Eqs. \rqn{3.54} and \rqn{3.56} \cite{note3}.

 The cases \rqn{3.54} and \rqn{3.56} are of
interest by themselves and are also of relevance below.
 In the both cases
${\cal U}=\mbox{diag}(\tau_x^2,\tau_x^2,\tau_z^2)$, yielding, in view
of Eq. \rqn{3.11},
 \be
S_+=2\max\left\{\sqrt{2}\tau_x, \sqrt{\tau_x^2+\tau_z^2}\right\},
 \e{3.53}
i.e., $S_+=2\sqrt{2}\tau_x$ for $\tau_x\ge\tau_z$ and
$S_+=2\sqrt{\tau_x^2+\tau_z^2}$ for $\tau_x\le\tau_z$.
 However, the optimal detector configurations are different in the
cases \rqn{3.54} and \rqn{3.56}, as follows.

\subsubsection{Case \protect\rqn{3.54}}
\label{IIIG1}

 In the case \rqn{3.54} ${\cal V}=\mbox{diag}(1,1,-1)$.
If $\tau_x\ge\tau_z$, $u_1=u_2=\tau_x^2$, and we can choose
$\vec{c}_1=\vec{x}$ and $\vec{c}_2=\vec{y}$, where $\vec{x},\
\vec{y}$, and $\vec{z}$ denote the unit vector along the
corresponding axis.
 Now $\vec{e}_1$ ($\vec{e}_2$) coincides with
$\vec{c}_1$ ($\vec{c}_2$), yielding $\delta=\nu=\phi$, where $\phi$
is the polar coordinate in the horizontal ($xy$) plane.
 The optimal configurations lie in the horizontal plane, being given
by
 \be
(\phi_a,\phi_a',\phi_b,\phi_b')= \pm(0,\pi/2,\pi/4,3\pi/4)+C,
 \e{3.19}
where $C$ is an arbitrary real number.
 Equation \rqn{3.19} follows from Eq. \rqn{3.13}, where we introduced the
double sign to take into account the both signs in Eq. \rqn{3.22}.
 Hence, Eq. \rqn{3.19} describes all
optimal configurations in the horizontal plane.

If $\tau_x\le\tau_z$, we choose
 \be
u_1=\tau_z^2,\quad u_2=u_3=\tau_x^2.
 \e{3.57}
Then $\vec{c}_1=-\vec{e}_1=\vec{z}$.
 Since $u_2=u_3$ (see Sec. \ref{IIIF}\,d), we choose $\vec{c}_2=
\vec{e}_2= \vec{x}\cos\phi_0+\vec{y}\sin\phi_0$, where $\phi_0$ is an
arbitrary number which equals the polar angle of $\vec{c}_2$ in the
$xy$ plane.
 The above expressions for $\vec{c}_1,\ \vec{c}_2,\
\vec{e}_1$, and $\vec{e}_2$ imply
 \be
\nu=\pi-\delta=\theta,
 \e{3.58}
where $\theta$ is the polar angle in a vertical plane (a plane
passing through the $z$ axis); $\theta$ is counted from the $z$ axis
in the direction where $\theta=\pi/2$ corresponds to $\vec{c}_2$.
 The polar coordinates of the optimal detector axes
$\theta_a,\ \theta_a',\ \theta_b$, and $\theta_b'$ as functions of
the parameter $\eta$ can be obtained from Eqs. \rqn{3.12},
\rqn{3.15}, \rqn{3.57}, and \rqn{3.58}.
 Thus, now optimal configurations lie in any vertical plane; they are
characterized by two continuous parameters, $\eta$ and $\phi_0$, in
agreement with Sec. \ref{IIIF}\,d.

As discussed above, the optimal configurations are horizontal for
$\tau_x>\tau_z$ and vertical for $\tau_x<\tau_z$.
 For $\tau_x=\tau_z$ we have the completely degenerate case (Sec.
\ref{IIIF}\,c) with $R=-{\cal V}=\mbox{diag}(-1,-1,1)$ which
corresponds to the rotation by $\pi$ around the $z$ axis.
 This rotation of qubit $a$ yields
$|\Psi_-\rangle\rightarrow|\Psi_+\rangle$.
 Hence, in this case the optimal configurations are given by those
for the Bell state $|\Psi_+\rangle$ \rqn{28} (see Sec.
\ref{IIIF}\,b).
 In particular, Eq. \rqn{3.19} provides the horizontal optimal
configurations \cite{kof08} for $|\Psi_+\rangle$, whereas the
vertical optimal configurations for $|\Psi_+\rangle$ follow from Eqs.
\rqn{3.13} and \rqn{3.58} and can be cast as \cite{kof08}
 \be
(\theta_a,\theta_a')= \pm(0,\pi/2)-C,\ \ (\theta_b,\theta_b')=
\pm(3\pi/4,\pi/4)+C.
 \e{3.60}
Here we introduced the double sign as in Eq. \rqn{3.19}.

\subsubsection{Case \protect\rqn{3.56}}
\label{IIIG2}

 In the case \rqn{3.56} ${\cal V}=\mbox{diag}(1,-1,1)$.
If $\tau_x\ge\tau_z$, $u_1=u_2=\tau_x^2$, and we can choose
$\vec{c}_1=\vec{e}_1=\vec{x}$ and $\vec{c}_2=-\vec{e}_2=\vec{y}$,
yielding $\nu=-\delta=\phi$.
 In this case the optimal configurations lie in the horizontal plane
and are given by Eq. \rqn{3.19} with $\phi_a\rightarrow-\phi_a$ and
$\phi_a'\rightarrow-\phi_a'$, i.e., by
 \be
(\phi_a,\phi_a')= \pm(0,-\pi/2)-C,\ \ (\phi_b,\phi_b')=
\pm(\pi/4,3\pi/4)+C.
 \e{3.61}

For $\tau_x\ge\tau_z$ Eq. \rqn{3.57} holds.
 As above, $\vec{c}_1=\vec{e}_1=\vec{z}$ and
$\vec{c}_2=\vec{x}\cos\phi_0+\vec{y}\sin\phi_0$, but
$\vec{e}_2=\vec{x}\cos\phi_0-\vec{y}\sin\phi_0$.
 Now the optimal detector axes for qubits $a$ and $b$ lie generally
in different vertical planes characterized, respectively, by the
polar coordinates $\phi_a$ and $\phi_b$ (in the $xy$ plane) of
$\vec{c}_2$ and $\vec{e}_2$, respectively, such that
 \be
\phi_b=-\phi_a=\phi_0.
 \e{3.68}
For a given value of $\phi_0$, the optimal configurations are given
by Eq. \rqn{3.13}.
 They are planar in two cases.
 For $\phi_0=0$ and $\pi$ the optimal configurations lie in the $xz$
plane and satisfy
 \be
\nu=\delta=\theta,
 \e{3.59}
whereas for $\phi_0=\pi/2$ and $3\pi/2$ they lie in the $yz$ plane
and satisfy
 \be
\nu=-\delta=\theta.
 \e{3.62}
In Eqs. \rqn{3.59} and \rqn{3.62} $\theta$ is counted from $\vec{z}$
in the direction where $\theta=\pi/2$ corresponds to $\vec{x}$ and
$\vec{y}$, respectively.

In the intermediate case $\tau_x=\tau_z$ the optimal configurations
coincide with those for the maximally entangled state characterized
by $R=-{\cal V}=\mbox{diag}(-1,1,-1)$ (see Sec. \ref{IIIF}\,b).
 This state, obtained from the singlet by the $\pi$ rotation of qubit
$a$ around the $y$ axis, is $|\Phi_+\rangle$ [Eq. \rqn{3.23}].
 In particular, the horizontal optimal configurations for
$|\Phi_+\rangle$ are given by Eq. \rqn{3.61}.
 The vertical optimal configurations are given by Eq. \rqn{3.13},
 \be
(\delta_a,\delta_a',\nu_b,\nu_b')= \pm(0,\pi/2,\pi/4,3\pi/4)+C.
 \e{3.63}
[The double sign is introduced here, as in Eq. \rqn{3.19}, to
describe all optimal configurations.]
 In the planar cases Eq. \rqn{3.63} simplifies according to Eqs.
\rqn{3.59} and \rqn{3.62}.

Note that Eq. \rqn{3.56} follows from Eq. \rqn{3.54} under the $\pi$
rotation of qubit $a$ around the $x$ axis.
 As a result, the above optimal configurations for the case \rqn{3.56}
can be obtained from those discussed for the case \rqn{3.54} by the
$\pi$ rotation of the detector axes for qubit $a$ around the $x$
axis.

\subsubsection{Nonmaximally entangled pure states}
\label{IIIG3}

As a simple application of the above results, consider nonmaximally
entangled pure states.
 Let us begin with the ``odd'' state
 \be
|\Psi\rangle=\cos\beta|10\rangle+\sin\beta|01\rangle
\quad(0<\beta<\pi/2).
 \e{4.16}
This state [as well as the states \rqn{2.4} and \rqn{4.7} below] is
relevant for experiments with superconducting phase qubits
\cite{mcd05,ste06}.
 For Eq. \rqn{4.16} ${\cal T}$ is given by Eq. \rqn{3.54} with
$\tau_x=\sin2\beta$ and $\tau_z=1$.
 Now $\tau_x\le\tau_z$, yielding \cite{gis91,pop92a}
 \be
S_+=2\sqrt{1+\sin^22\beta}.
 \e{3.72}
 For $\beta=\pi/4$ the state \rqn{4.16} is maximally entangled and
coincides with the Bell state $|\Psi_+\rangle$ (see Sec.
\ref{IIIG1}).
 For the nonmaximally entangled state \rqn{4.16} ($\beta\ne\pi/4$) the
optimal configurations lie in a vertical plane and depend on two
continuous parameters, as discussed in Sec. \ref{IIIG1}, where now
$u_1=1$ and $u_2=\sin^22\beta$.

Any entangled pure two-qubit state can be obtained from the state
\rqn{4.16} by local rotations $R_a$ and $R_b$ of the qubits $a$ and
$b$, respectively.
 These rotations satisfy ${\cal T}'=R_a{\cal
T}R_b^T$, where ${\cal T}'$ and ${\cal
T}=\mbox{diag}(\sin2\beta,\sin2\beta,-1)$ are the matrices \rqn{3.17}
for the given state and Eq. \rqn{4.16}, respectively.
 In view of the invariance of $S$ under local rotations \rqn{3.16}, we
obtain that for an arbitrary entangled pure state, $S_+$ is given by
Eq. \rqn{3.72}, whereas the optimal configurations coincide with
those for the state \rqn{4.16} with the substitutions \rqn{3.16b}.

In particular, consider the ``even'' state
 \be
|\Phi\rangle=\cos\beta|11\rangle+\sin\beta|00\rangle
\quad(0<\beta<\pi/2).
 \e{5.12}
It satisfies Eq. \rqn{3.56} and has the same $\tau_x$, $\tau_z$, and
$S_+$ as the state \rqn{4.16}.
 For $\beta=\pi/4$ the state \rqn{5.12} coincides with
$|\Phi_+\rangle$ (see Sec. \ref{IIIG2}).
 For $\beta\ne\pi/4$ the optimal configurations are vertical and are
obtained as discussed in Sec. \ref{IIIG2}, using $u_1=1$ and
$u_2=\sin^22\beta$.
 These configurations can be obtained from the optimal configurations
for the state \rqn{4.16} by the $\pi$ rotation of the detector axes
for qubit $a$ around the $x$ axis (cf. the last paragraph in Sec.
\ref{IIIG2}).

\section{Effects of decoherence}
\label{IV}

\subsection{Description of decoherence}

To investigate effects of decoherence on the BI violation, we assume
the following simplified picture of the experiment: after a fast
preparation of the initial state $\rho_0$, the qubits undergo
decoherence during time $t$ resulting in the state $\rho$, then a
fast measurement follows.
 Now in Eq. \rqn{3.5}
 \be
\rho={\cal L}(\rho_0),
 \e{5.1}
where the superoperator (linear map) $\cal L$ describes decoherence
of the qubit pair.
 We assume independent (local) decoherence of each qubit and the
absence of any other evolution, so that
 \be
{\cal L}={\cal L}_a\otimes{\cal L}_b.
 \e{5.2}

We consider Markovian decoherence which involves energy relaxation at
the zero temperature (i.e., spontaneous transitions
$|1\rangle\rightarrow|0\rangle$) and pure dephasing.
 The assumption of the zero temperature, $T=0$, simplifies the
formulas below.
 It is applicable to low-temperature systems (such as superconducting
phase qubits), with $k_BT\ll E_q$, where $k_B$ is the Boltzmann
constant and $E_q$ is the energy separation of the qubit.
 Moreover, the BI violation conditions were found in Ref.
\onlinecite{kof} to depend very weakly on the temperature.
 Under the above assumptions, the elements of the density matrix
$\rho_k(t)$ of qubit $k\ (k=a,b)$ obey the Bloch equations
\cite{coh92}
 \bea
&&\dot{\rho}_{11}^k(t)=-\dot{\rho}_{00}^k(t)=-\rho_{11}^k(t)/T_1^k,
\nonumber\\
&&\dot{\rho}_{10}^k(t)=-\rho_{10}^k(t)/T_2^k,\quad
\dot{\rho}_{01}^k(t)=-\rho_{01}^k(t)/T_2^k.
 \ea{4.1}
Here $T_1^k$ and $T_2^k$ are the decoherence times, obeying
$T_2^k\le2T_1^k$, where the inequality occurs in the presence of
(pure) dephasing which proceeds with the rate
$\Gamma_d^k=1/T_2^k-1/(2T_1^k)$.
 In the derivation of Eqs. \rqn{4.1} it is usually assumed that
$E_q/\hbar\gg1/T_1^k,1/T_2^k$.
 This condition is satisfied in many systems.
 In particular, it holds for superconducting phase qubits.

Equations \rqn{4.1} can be easily solved, providing the linear map
$\rho_k(t)={\cal L}_k(\rho_k(0))$.
 The superoperator ${\cal L}_k$ can be written in terms of the Kraus
operators (the operator-product form \cite{nie00}),
 \be
{\cal L}_k(\rho_k(0))=\sum_{i=1}^3K_i^k\rho_k(0)(K_i^k)^\dagger,
 \e{5.3}
where the Kraus operators are the special case for $T=0$ of the Kraus
operators obtained in Ref. \onlinecite{kof},
 \bea
&&K_1^k=\left(\begin{array}{cc}
0&0\\
\sqrt{1-\gamma_k}&0
\end{array}\right),\ \
 K_2^k=\left(\begin{array}{cc}
\sqrt{\gamma_k}&0\\
0&\mu_k
\end{array}\right),\nonumber\\
&&K_3^k=\left(\begin{array}{cc}
0&0\\
0&\sqrt{1-\mu_k^2}
\end{array}\right).
 \ea{5.8}
Here
 \be
\gamma_k=e^{-t/T_1^k},\
\mu_k=\lambda_k/\sqrt{\gamma_k}=e^{-\Gamma_d^kt},\
\lambda_k=e^{-t/T_2^k}.
 \e{4.2}
 The Kraus operators \rqn{5.8} take into account energy relaxation
and dephasing simultaneously, extending thus the previously known
Kraus operators \cite{nie00} which describe either effect separately.

Combining Eqs. \rqn{5.1}, \rqn{5.2}, and \rqn{5.3} yields the
expression for the two-qubit superoperator
 \be
\rho={\cal L}(\rho_0)=\sum_{i,j=1}^3K_{ij}\rho_0K_{ij}^\dagger,\quad
K_{ij} = K_i^a\otimes K_j^b
 \e{4.6}
through the nine two-qubit Kraus operators $K_{ij}$.
 As a result of decoherence, the initial two-qubit density matrix
$\rho_0=\{\rho_{ij}^0\}$ evolves after time $t$ to
 \begin{widetext}
 \be
\rho=\left(\begin{array}{cccc}
\gamma_a\gamma_b\rho_{11}^0&\gamma_a\lambda_b\rho_{12}^0
&\lambda_a\gamma_b\rho_{13}^0&\lambda_a\lambda_b\rho_{14}^0\\
\gamma_a\lambda_b\rho_{21}^0
&\gamma_a(\gamma_b'\rho_{11}^0+\rho_{22}^0)
&\lambda_a\lambda_b\rho_{23}^0
&\lambda_a(\gamma_b'\rho_{12}^0+\rho_{24}^0)\\
\lambda_a\gamma_b\rho_{31}^0&\lambda_a\lambda_b\rho_{32}^0
&\gamma_b(\gamma_a'\rho_{11}^0+\rho_{33}^0)
&\lambda_b(\gamma_a'\rho_{13}^0+\rho_{34}^0)\\
\lambda_a\lambda_b\rho_{41}^0
&\lambda_a(\gamma_b'\rho_{21}^0+\rho_{42}^0)
&\lambda_b(\gamma_a'\rho_{31}^0+\rho_{43}^0)
&\gamma_a'\gamma_b'\rho_{11}^0+\gamma_a'\rho_{22}^0+
\gamma_b'\rho_{33}^0+\rho_{44}^0
\end{array}\right),
 \e{4.3}
 \end{widetext}
where $\gamma_{a(b)}'=1-\gamma_{a(b)}$.
 The ordering of the elements of the density matrix is specified
after Eq. \rqn{3.69}.

 Decoherence generally breaks the invariance of $S$ with respect to
local transformations of $\rho_0$ and the corresponding rotations of
the detectors [cf. Eq. \rqn{3.16}].
 Aa a result, locally equivalent initial states may yield different
maximal violations of the BI.
 However, in the present model of decoherence $S$ is still invariant
under local rotations of the initial state around the $z$ axis and
the corresponding rotations of the detectors.

\subsection{Bell operator modified by decoherence}
\label{IVB}

It is useful to recast $S={\rm Tr}({\cal B}\rho)$ in the form $S={\rm
Tr}({\cal B}_d\rho_0)$ with the modified Bell operator ${\cal
B}_d=({\cal L}_a^*\otimes{\cal L}_b^*)({\cal B})$ or
 \be
{\cal B}_d=A_dB_d-A_dB_d'+A_d'B_d+A_d'B_d'.
 \e{5.25}
Here ${\cal L}_k^*\ (k=a,b)$ is the map adjoint (dual) to ${\cal
L}_k$ that moves observables of the quantum system \cite{ali87} and
 \be
A_d={\cal L}_a^*(A)=\sum_{i=1}^3(K_i^a)^\dagger AK_i^a,
 \e{5.4}
etc.
 From Eqs. \rqn{3.24}, \rqn{5.8}, and \rqn{5.4} we obtain
 \be
A_d=(\gamma_a-1)a_z+\vec{q}_a\cdot\vec{\sigma}_a,\ \
\vec{q}_a=(\lambda_aa_x,\lambda_aa_y,\gamma_aa_z),
 \e{5.5}
where $\lambda_k$ is defined in Eq. \rqn{4.2}.
 Expressions similar to Eq. \rqn{5.5} hold also for $A_d',\
B_d$, and $B_d'$.
 The maximal violation of the BI occurs always for a pure initial
state (an eigenvector of the Hermitian operator ${\cal B}_d$
corresponding to the maximal eigenvalue).

We performed numerical calculations of the maximum $S_{\rm max}$ of
$S$ over all the states and observation directions as a function of
the decoherence parameters with the help of the Mathematica 6 routine
NMaximize.
 We used two methods.
 The first method is based on the fact that $S_{\rm max}$ is equal to
the maximum of the greatest eigenvalue of ${\cal B}_d$ over the
directions $\vec{a}$, $\vec{a}'$, $\vec{b}$, and $\vec{b}'$, the
optimal state being given by the corresponding eigenvector.
 The detector axes are determined by eight independent parameters.
 However, due to the invariance of $S$ with respect to rotations
around the $z$ axis, we can reduce the number of the fitting
parameters from eight to six by setting, say, $a_y=b_y=0$.
 Combining the above invariance with the freedom of rotations of the
detectors (see Sec. \ref{III}, especially Fig. \ref{f1}) yields that
one of the detectors can be fixed in any position, say, along the $x$
axis, so that, e.g., $\vec{a}=(1,0,0)$.
 This further reduces the number of the fitting parameters to five.
 The second method involves the analytical approach of Sec.
\ref{III}, as discussed in Sec. \ref{IVC2}.

\subsection{States maximizing the BI violation}
\label{IVC2}

 Our numerical calculations by the method of Sec. \ref{IVB} show that
$S_{\rm max}$ can be always obtained for a pure state
$\rho_0=|\Psi\rangle\langle\Psi|$ of the form
 \be
|\Psi\rangle=c_1|11\rangle+c_2|10\rangle+c_3|01\rangle+c_4|00\rangle,
 \e{5.7}
where $c_i$ are {\em real} coefficients such that
$\sum_{i=1}^4c_i^2=1$.
 Note that the state \rqn{5.7} is not the most general two-qubit state,
even if one takes into account the invariance of $S$ with regard to
local $z$ rotations.
 Indeed, the general two-qubit state depends on six independent
parameters, and local $z$ rotations can decrease this number to four,
whereas Eq. \rqn{5.7} depends on three parameters.

If the state \rqn{5.7} with certain values of $c_i$ provides $S_{\rm
max}$ for some choice of the decoherence parameters, so does the
state $|\Psi(\alpha_a,\alpha_b)\rangle=U_z(\alpha_a)\otimes
U_z(\alpha_b)|\Psi\rangle$, where $\alpha_a$ and $\alpha_b$ are any
real numbers and $U_z(\alpha)=e^{-i\alpha\sigma_z/2}$ rotates a qubit
around the $z$ axis by angle $\alpha$.
 With the accuracy to an overall phase factor,
 \bea
&|\Psi(\alpha_a,\alpha_b)\rangle=&c_1|11\rangle
+c_2e^{i\alpha_b}|10\rangle+
c_3e^{i\alpha_a}|01\rangle\nonumber\\
&&+e^{i(\alpha_a+\alpha_b)}c_4|00\rangle.
 \ea{4.30}
All optimal detector configurations for this state are obtained from
those for the state \rqn{5.7} by rotating the detectors for qubits
$a$ and $b$ around the $z$ axis by the angles $\alpha_a$ and
$\alpha_b$, respectively.

It is convenient for numerical calculations to express the
coefficients $c_i$ through the parameters $\kappa_1,\ \kappa_2$, and
$\kappa_3$ by
\begin{widetext}
 \be
(c_1,c_2,c_3,c_4)= (\sin\kappa_1,\cos\kappa_1\sin\kappa_2,
\cos\kappa_1\cos\kappa_2\sin\kappa_3,
\cos\kappa_1\cos\kappa_2\cos\kappa_3).
 \e{5.11}
Now in Eq. \rqn{4.3} the elements $\rho_{ij}^0=c_ic_j$ are real and
we obtain from Eq. \rqn{3.17} the matrix
 \be
{\cal T}=2\left(\begin{array}{ccc}
\lambda_a\lambda_b(\rho_{23}^0+\rho_{14}^0)&0&\lambda_a[\rho_{24}^0+
(1-2\gamma_b)\rho_{13}^0]\\
0&\lambda_a\lambda_b(\rho_{23}^0-\rho_{14}^0)&0\\
\lambda_b[\rho_{34}^0+(1-2\gamma_a)\rho_{12}^0]&0&1/2-d\rho_{11}^0
-\gamma_a\rho_{22}^0 -\gamma_b\rho_{33}^0
\end{array}\right),
 \e{5.10}
\end{widetext}
where $d=\gamma_a+\gamma_b-2\gamma_a\gamma_b$ (note that $0\le
d\le1$).
 In the matrix ${\cal T}$ \rqn{5.10} only the $xz$ and $zx$
off-diagonal elements are nonvanishing.
 This allows us to obtain an analytical solution for $S_+$ and the
optimal configurations, using the formalism of Sec. \ref{III}, as
follows.

From Eq. \rqn{5.10} we obtain that the nonzero elements of ${\cal
U}={\cal T}^T{\cal T}$ equal
 \bea
&{\cal U}_{xx}=&4\lambda_b^2\{\lambda_a^2(\rho_{14}^0+\rho_{23}^0)^2
+[\rho_{34}^0+(1-2\gamma_a)\rho_{12}^0]^2\},\nonumber\\
&{\cal U}_{yy}=&4\lambda_a^2\lambda_b^2(\rho_{14}^0-\rho_{23}^0)^2,
\nonumber\\
&{\cal U}_{zz}=&4\lambda_a^2[\rho_{24}^0+(1-2\gamma_b)\rho_{13}^0]^2
+g^2,
\nonumber\\
&{\cal U}_{xz}=&{\cal U}_{zx}=4\lambda_a^2\lambda_b(\rho_{14}^0
+\rho_{23}^0)[\rho_{24}^0+(1-2\gamma_b)\rho_{13}^0]\nonumber\\
&&-2\lambda_b[\rho_{34}^0+(1-2\gamma_a)\rho_{12}^0]g,
 \ea{a1}
where $g=1-2d\rho_{11}^0-2\gamma_a\rho_{22}^0-2\gamma_b\rho_{33}^0$.
 It is easy to see that the eigenvalues of ${\cal U}$ equal
${\cal U}_{yy}$ and
 \be
U_\pm=({\cal U}_{xx}+{\cal U}_{zz})/2 \pm\sqrt{({\cal U}_{xx}-{\cal
U}_{zz})^2/4+{\cal U}_{xz}^2},
 \e{a2}
the corresponding eigenvectors being $\vec{y}$ and
 \be
\vec{c}_\pm=[(U_\pm-{\cal U}_{xx})^2+{\cal U}_{xz}^2]^{-1/2} ({\cal
U}_{xz},0,U_\pm-{\cal U}_{xx}).
 \e{a3}
Thus, we obtain that
 \be
u_1=U_+,\quad u_2=\max\{U_-,{\cal U}_{yy}\},
 \e{a4}
whereas $\vec{c}_1=\vec{c}_+$ and $\vec{c}_2=\vec{c}_-$ if $u_2=U_-$
or $\vec{c}_2=\vec{y}$ if $u_2={\cal U}_{yy}$.
 These values of $u_1,\ u_2,\ \vec{c}_1$, and $\vec{c}_2$ provide
$S_+$ and the optimal detector configurations for the state
\rqn{5.7}, on employing Eqs. \rqn{3.11}, \rqn{3.74}, \rqn{5.10}, and
\rqn{5.22}-\rqn{3.7} [or \rqn{3.22}, \rqn{3.12}, and \rqn{3.15}].
 Generally, the optimal detector axes lie in the $xz$ plane, when
$\vec{c}_2=\vec{c}_-$, or in two planes passing through the $\vec{y}$
axes, when $\vec{c}_2=\vec{y}$ (since then $\vec{e}_2=\vec{y}$).

We use this solution to obtain $S_{\rm max}$ and the corresponding
optimal state, by maximizing $S_+$ numerically over the parameters
$\kappa_1,\ \kappa_2$, and $\kappa_3$.
 This procedure is significantly faster than the numerical
method described in Sec. \ref{IVB}.
 Before the discussion of the results of numerical calculations in
Sec. \ref{IVD}, we consider important cases which admit simple
analytical solutions.

\subsection{Analytical solutions}

\subsubsection{Horizontal optimal configurations}

An especially simple solution is obtained when the optimal
observation axes lie in the $xy$ (horizontal) plane.
 Then Eqs. \rqn{5.25} and \rqn{5.5} with $a_z=0$ yield
${\cal B}_d=\lambda_a\lambda_b{\cal B}$, and hence
$S=\lambda_a\lambda_bS_0$, where $S_0={\rm Tr}({\cal B}\rho_0)$ is
obtained in the absence of decoherence (Sec. \ref{II}).
 As a result, the value of $S$ maximized over all states and
horizontal observation directions is
 \be
S_h=2\sqrt{2}\lambda_a\lambda_b.
 \e{5.6}

This value is obtained only for the maximally entangled states which
have horizontal optimal detector configurations in the ideal case,
as, e.g., the states $|\Psi_+\rangle$ and $|\Phi_+\rangle$ discussed
in Sec. \ref{IIIG}.
 All such states are given by the expressions
 \bea
&&|\Psi\rangle=(|10\rangle+e^{i\alpha}|01\rangle)/\sqrt{2},
\label{2.3}\\
&&|\Phi\rangle=(|11\rangle+e^{i\alpha}|00\rangle)/\sqrt{2}.
 \ea{2.4}
Indeed, taking into account that $S$ is invariant under identical
rotations of qubits and detectors, Eq. \rqn{3.16}, and that all
maximally entangled states are related by a rotation of one of the
qubits (cf. Sec. \ref{IIIF}\,b), we obtain that all maximally
entangled states with horizontal optimal configurations result from
the state $|\Psi_+\rangle$ (or $|\Phi_+\rangle$) on applying to one
of the qubits an arbitrary rotation around the $z$ axis and possibly
a $\pi$ rotation around the $x$ axis, since only such rotations do
not take the detector axes out of the horizontal plane.
 All the resulting states are given by Eqs. \rqn{2.3} and \rqn{2.4}.

The states \rqn{2.3} and \rqn{2.4} are special cases of the states
\rqn{4.7} and \rqn{4.10}, which are discussed in detail in Sec.
\ref{IVD2}.
 In particular, in Sec. \ref{IVD2} the validity conditions of Eq.
\rqn{5.6} are obtained.

\subsubsection{Even and odd states}
\label{IVD2}

It is of interest to consider maximal violations of the BI for the
special cases of Eq. \rqn{5.7}, the classes of ``odd'' and ``even''
nonmaximally entangled states, Eqs. \rqn{4.16} and \rqn{5.12},
respectively.
 For the odd and even states, respectively, Eq. \rqn{5.10} yields
 \be
{\cal T}=\mbox{diag}(\lambda_a\lambda_b\sin2\beta,
\lambda_a\lambda_b\sin2\beta,1-\gamma_+-\gamma_-\cos2\beta),
 \e{4.4}
 \be
{\cal T}=\mbox{diag}(\lambda_a\lambda_b\sin2\beta,
-\lambda_a\lambda_b\sin2\beta,1-d-d\cos2\beta),
 \e{4.5}
where $\gamma_\pm=\gamma_a\pm\gamma_b$.
 Equations \rqn{4.4} and \rqn{4.5} have the form of Eqs. \rqn{3.54} and
\rqn{3.56}, respectively, with $\tau_x=\lambda_a\lambda_b\sin2\beta$,
whereas $\tau_z=\gamma_++\gamma_-\cos2\beta-1$ and
$\tau_z=1-d-d\cos2\beta$ for the odd and even states, respectively.
 Only positive values of the above expressions for $\tau_z$ are of
interest for the BI violation, since for $\tau_z\le0$ $\det({\cal
T})\ge0$, and the BI violation is impossible (see Sec. \ref{IIIB}).

For the both states we obtain \cite{note2}, in view of Eq.
\rqn{3.53},
 \be
S_+=2\max\left\{\sqrt{2}\lambda_a\lambda_b\sin2\beta,
\sqrt{\lambda_a^2\lambda_b^2\sin^22\beta+\tau_z^2}\right\}.
 \e{5.15}
 Correspondingly, the optimal configurations for the odd and even
states (with $\tau_z>0$) are given in Secs. \ref{IIIG1} and
\ref{IIIG2}, respectively.
 (The optimal configurations for the case $\tau_z\le0$
can be obtained, if necessary, from the results of Sec. \ref{III}.)
 In particular, the optimal configurations lie in the horizontal (a
vertical) plane if the first (second) term in the braces in Eq.
\rqn{5.15} is greater than the other term.

According to Eq. \rqn{5.15}, for a given odd or even state the value
of $S_+$ depends on $\beta$.
 In particular, when the first term in the braces in Eq. \rqn{5.15}
exceeds the second one, $S_+=2\sqrt{2}\lambda_a\lambda_b\sin2\beta$.
 This holds only when both $\lambda_a\lambda_b$ and $\beta$ are
sufficiently large, which implies relatively weak pure dephasing
\cite{kof}.
 We do not dwell here upon the case of an arbitrary $\beta$, since
we are interested in a state which maximizes $S_+$.

In the ideal case, $S_{\rm max}$ is obtained for a maximally
entangled state \cite{pop92}, but in the presence of decoherence this
is not necessarily so.
 It is still of interest to consider $S_+$ for the
maximally entangled odd and even states ($\beta=\pi/4$), i.e.,
$|\Psi_+\rangle$ and $|\Phi_+\rangle$, respectively.
 For $|\Psi_+\rangle$ and $|\Phi_+\rangle$ Eq. \rqn{5.15} with
$\beta=\pi/4$ becomes, respectively,
 \be
S_+=S_+^\Psi=2[\lambda_a^2\lambda_b^2+ \max\{\lambda_a^2\lambda_b^2,
(\gamma_+-1)^2\}]^{1/2},
 \e{5.19}
 \be
S_+=S_+^\Phi=2[\lambda_a^2\lambda_b^2+\max\{\lambda_a^2\lambda_b^2,
(1-d)^2\}]^{1/2}.
 \e{5.20}
When the first term in the braces in Eqs. \rqn{5.19} and \rqn{5.20}
is greater than the second term, Eqs. \rqn{5.19} and \rqn{5.20}
reduce to Eq. \rqn{5.6}.
 One can show that
$1-d\ge|\gamma_+-1|$, the equality being for $\gamma_a=\gamma_b$
equal to 1 or 0.
 This yields that $S_+^\Phi>S_+^\Psi$ when the second term in the
braces in Eqs. \rqn{5.19} and \rqn{5.20} is greater than the first
term and $t>0$.
 Hence, in the presence of decoherence, the BI violation for
$|\Phi_+\rangle$ is generally greater than for $|\Psi_+\rangle$.
 This means that the BI violation duration $\tau_B$, defined by
$S_+(t=\tau_B)=2$, is generally greater for $|\Phi_+\rangle$ than for
$|\Psi_+\rangle$.

Generally, maximally entangled states are not optimal for the BI
violations, unless the case \rqn{5.6} is optimal.
 Consider the maximal BI violations for the classes of the even and
odd states.
 To this end, $S_+$ in Eq. \rqn{5.15} should be maximized with respect
to $\beta$.
 This yields for the odd state \rqn{4.16}
 \bes{5.13}
 \bea
&S_+=&2\lambda_a\lambda_b[1+\max\{1,
(\gamma_+-1)^2/(\lambda_a^2\lambda_b^2-\gamma_-^2)\}]^{1/2}
\nonumber\\
&&\mbox{if } \lambda_a^2\lambda_b^2>d_1,\label{5.13a}\\
&S_+=&2\max\{\sqrt{2}\lambda_a\lambda_b, d_2\}\ \ \mbox{if }
\lambda_a^2\lambda_b^2\le d_1,
 \ea{5.13b}
 \ese
where $d_1=\gamma_-^2+|\gamma_-|(\gamma_+-1)\le1$ and
$d_2=|\gamma_++|\gamma_-|-1|\le1$, and for the even state \rqn{5.12}
 \bes{5.16}
 \be
S_+=2\lambda_a\lambda_b\left[1+\max\left\{1,
\frac{(1-d)^2}{\lambda_a^2\lambda_b^2-d^2} \right\}\right]^{1/2}\ \
\mbox{if } \lambda_a^2\lambda_b^2>d,
 \e{5.16a}
 \be
S_+=2\quad\mbox{if } \lambda_a^2\lambda_b^2\le d.
 \e{5.16b}
 \ese
Equations \rqn{5.13a} and \rqn{5.16a}, in contrast to Eqs.
\rqn{5.13b} and \rqn{5.16b}, can describe a violation of the BI.
 When the first term in the braces in Eq. \rqn{5.13a} or \rqn{5.16a} is
greater than the second term, the case \rqn{5.6} is realized and the
corresponding state is maximally entangled ($\beta=\pi/4$).
 In the opposite case, the optimal odd and even states generally are
not maximally entangled, being characterized by the following values
of $\beta$, respectively,
 \bea
&&\beta=\arccos[\gamma_-(\gamma_+-1)/(\lambda_a^2\lambda_b^2
-\gamma_-^2)]/2,\label{4.17}\\
&& \beta=\pi/2-\arccos[(d-d^2)/(\lambda_a^2\lambda_b^2-d^2)]/2.
 \ea{5.26}
Note that Eq. \rqn{5.13b}, with the second term greater than the
first term, and Eq. \rqn{5.16b} are obtained for a nonentangled
initial state: $|10\rangle$ if $T_1^a<T_1^b$ or $|10\rangle$ if
$T_1^a>T_1^b$ for Eq. \rqn{5.13b} and $|00\rangle$ for Eq.
\rqn{5.16b}.

Numerical calculations show that the maximal BI violation in the
optimal even state is greater than or equal to that obtained in the
optimal odd state.
 The differences in $S_+$ for the even and odd states can appear only
in the case when the detectors are in a vertical plane.
 In this case the optimal states are generally nonmaximally
entangled.
 The reason for this is that the observables whose axes do not lie
in the horizontal plane are sensitive to energy relaxation.
 As a result, for instance, in the optimal even state the amplitude
of $|11\rangle$ is less than the amplitude of $|00\rangle$, since
this bias reduces spontaneous decay and hence diminishes the
detrimental effect of relaxation on the BI violation.
 By the same reason, in the optimal odd state the amplitude of the
excited qubit with a smaller $\gamma_k$ (shorter $T_1^k$) is reduced.
 However, in the case $\gamma_a=\gamma_b$ the optimal odd state, in
contrast to the even state, is maximally entangled, and no relaxation
reduction is achieved.

The states \rqn{4.16} and \rqn{5.12} discussed here are special cases
of the general odd and even states, respectively ($0\le\alpha<2\pi,\
0<\beta<\pi/2$),
 \bea
&&|\Psi\rangle=\cos\beta|10\rangle+e^{i\alpha}\sin\beta|01\rangle,
\label{4.7}\\
&&|\Phi\rangle=\cos\beta|11\rangle+e^{i\alpha}\sin\beta|00\rangle.
 \ea{4.10}
The state \rqn{4.7} is relevant for experiments with superconducting
phase qubits \cite{mcd05,ste06}.
 The states \rqn{4.7} and \rqn{4.10} result from Eqs. \rqn{4.16} and
\rqn{5.12}, respectively, under the rotation of qubit $a$ around the
$z$ axis by the angle $\alpha$.
 Hence, the results for $S_+$ obtained in the present paper for the
odd and even states hold, respectively, also for the states \rqn{4.7}
and \rqn{4.10} with the same $\beta$, the corresponding optimal
configurations being modified by the rotation of the detectors for
qubit $a$ around the $z$ axis by $\alpha$ [see also the paragraph
containing Eq. \rqn{4.30}].

\subsection{Numerical results and discussion}
\label{IVD}

\subsubsection{Pure dephasing}

 First, let us discuss the case of the absence of energy relaxation,
$\gamma_a=\gamma_b=1$, when decoherence occurs due to pure dephasing.
 Our calculations show that now $|\Psi_+\rangle$ and $|\Phi_+\rangle$
are optimal states providing $S_{\rm max}$.
 In this case $\gamma_+=2$, $\gamma_-=d=0$ and both Eqs. \rqn{5.13}
and \rqn{5.16} yield \cite{sam03,vel03}
 \be
S_{\rm max}=2\sqrt{1+\lambda_a^2\lambda_b^2}.
 \e{5.24}
Thus, now the BI can be violated for any level of decoherence.
 The value \rqn{5.24} is achieved for the observation axes lying in a
vertical plane \cite{sam03}.
 They are described by the results of Secs. \ref{IIIG1} and
\ref{IIIG2}, taking into account that now ${\cal
T}=\mbox{diag}(\lambda_a\lambda_b, \pm\lambda_a\lambda_b,\mp1)$, the
upper (lower) sign correspomding to the odd (even) state.
 Note that the states \rqn{4.7} and \rqn{4.10} are also optimal.
 However, not all maximally entangled states are optimal now
\cite{vel03}, since $S$ is generally non-invariant with respect to
local transformations.

\subsubsection{Identical decoherence of the qubits}

Next, consider the case of
identical decoherence for the qubits of a pair.
 Now $\mu_a=\mu_b=\mu$ and
 \be
\gamma_a=\gamma_b=\gamma,\ \lambda_a=\lambda_b=\lambda,\
d=2\gamma(1-\gamma),\ \gamma_+=2\gamma.
 \e{4.13}
 The numerical calculations show that in this case the maximal $S$
can be always obtained with the even state \rqn{5.12}, i.e., $S_{\rm
max}$ is given by Eq. \rqn{5.16} with the account of Eq. \rqn{4.13},
 \bes{4.31}
 \be
S_{\rm max}=2\lambda^2\left[1+\max\left\{1,
\frac{(1-d)^2}{\lambda^4-d^2} \right\}\right]^{1/2}\ \ \mbox{if }
\lambda^4>d,
 \e{4.31a}
 \be
S_{\rm max}=2\quad\mbox{if } \lambda^4\le d.
 \e{4.31b}
 \ese
Now the optimal odd state is the Bell state $|\Psi_+\rangle$, so that
Eq. \rqn{5.13} reduces to Eq. \rqn{5.19} with the account of Eq.
\rqn{4.13},
 \be
S_+=S_+^\Psi=2\sqrt{\lambda^4+\max\{\lambda^4,(2\gamma-1)^2\}}.
 \e{4.32}

\begin{figure}[htb]
\includegraphics[width=8.5cm]{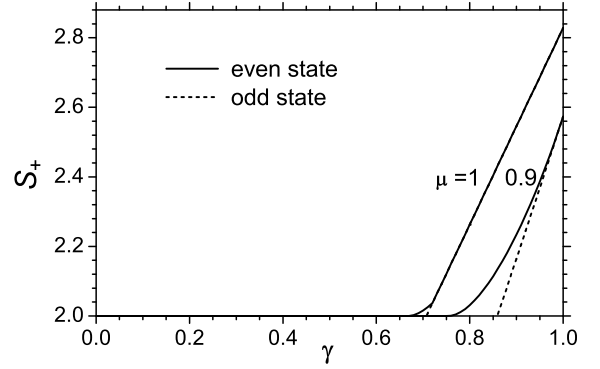}
\caption{$S_+$ versus the relaxation parameter $\gamma$ for $\mu=1$
(no pure dephasing) and $\mu=0.9$. Solid lines: $S_+=S_{\rm max}$
\rqn{4.31} with the even state \rqn{5.12}, dashed lines: Eq.
\rqn{4.32} for the odd state $|\Psi_+\rangle$ \rqn{28}.}
 \label{f2}\end{figure}

Figure \ref{f2} shows the dependence of $S_+$ on $\gamma$ with
$\mu=1$ (no pure dephasing) and $\mu=0.9$ for the even state (when
$S_+=S_{\rm max}$) and the odd state $|\Psi_+\rangle$.
 Note that the straight segments in Fig. \ref{f2} (in particular, the
both dashed lines) correspond to $S_+=S_h=2\sqrt{2}\lambda^2$ [Eq.
\rqn{5.6}].
 Equation \rqn{4.31a} and Fig. \ref{f2} imply that the violation
of the BI is possible only for $\gamma>2/3$, this limit being
approached in the absence of pure dephasing ($\mu=1$).
 As a result, for a given value of $T_1$, the maximal BI violation
duration $\tau_B$ is obtained for the even state when $\mu=1$, being
given by $\gamma=e^{-\tau_B/T_1}=2/3$ or
$\tau_B=T_1\ln1.5\approx0.405T_1$.
 For comparison, we mention that in the case $\mu=1$ the odd state
yields $S_+=S_h=2\sqrt{2}\gamma$, so that the BI can be violated only
for $\gamma>1/\sqrt{2}\approx0.707$.
 This corresponds to the longest $\tau_B$ for the odd state with a
given $T_1$ equal to \cite{jam06,kof}
$\tau_B=T_1\ln2/2\approx0.347T_1$.
 Note that, in contrast to Ref. \onlinecite{jam06}, we obtain
different values of $\tau_B$ for the even and odd states.

\begin{figure}[htb]
\includegraphics[width=8.5cm]{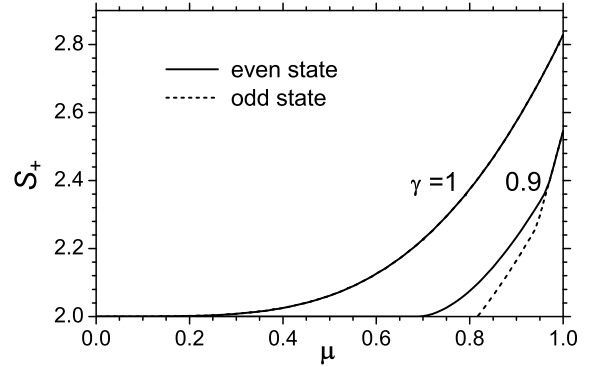}
\caption{$S_+$ versus the pure dephasing parameter $\mu$ for
$\gamma=1$ (no decay) and $\gamma=0.9$.
 Solid lines: $S_+=S_{\rm max}$ \rqn{4.31} with the even state
\rqn{5.12}, dashed lines: Eq. \rqn{4.32} for the odd state
$|\Psi_+\rangle$ \rqn{28}.}
 \label{f3}\end{figure}

Figure \ref{f3} shows the dependence of $S_+$ on the pure dephasing
parameter $\mu$.
 For $\gamma=1$ the plots for the odd and even state coincide and are
given by Eq. \rqn{5.24}, $S_+=S_{\rm
max}=\sqrt{1+\mu^4}=\sqrt{1+\lambda^4}$.
 In this case violations of
the BI can be achieved for any degree of pure dephasing.

\begin{figure}[htb]
\includegraphics[width=8.cm]{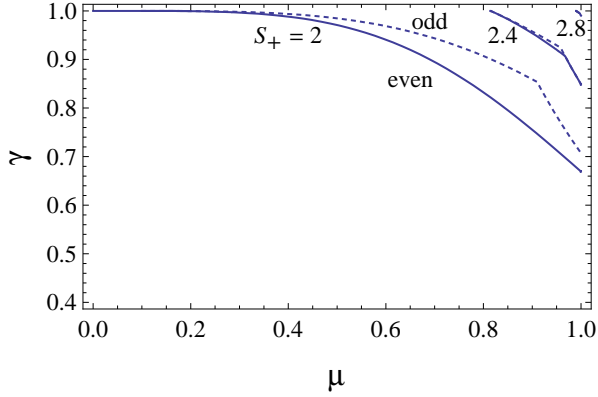}
\caption{Contour plot of the maximum $S_+$ of $S$ versus $\gamma$
and $\mu$.
  Solid lines: the even state \rqn{5.12} ($S_+=S_{\rm max}$), dashed
lines: the odd state $|\Psi_+\rangle$.}
 \label{f4}\end{figure}

Figure \ref{f4} is a contour plot of $S_+$ as a function of $\gamma$
and $\mu$.
 The boundary of the region of the BI violation is shown by the solid
line with $S_+=S_{\rm max}=2$.
 This boundary is obtained when the second term dominates in the
braces in Eq. \rqn{4.31a}.
 The kinks on the curves in Figs. \ref{f2}-\ref{f4} correspond to a
change of the dominating term in the braces in Eqs. \rqn{4.31a} and
\rqn{4.32}.
 Figures \ref{f2}-\ref{f4} show that $S_+$ for the odd state is
generally lower than $S_{\rm max}$, which results in more stringent
conditions on the decoherence parameters required for the BI
violation than for the even state.
 The difference is significant when $S_+-2$ is small.
 However, for $S_+\ge2.4$ there is practically no difference in the
values of $S_+$ for the odd and even states.

\subsubsection{No decoherence in one qubit}

 Consider the extreme case of nonequal
decoherence of the qubits, i.e., the case when decoherence is absent
in one of the qubits, e.g., in qubit $b$.
 Now $\gamma_b=\lambda_b=1$, $\gamma_+=1+\gamma_a,\
d=d_1=-\gamma_-=1-\gamma_a,\ d_2=1$, and hence Eqs. \rqn{5.13} and
\rqn{5.16} coincide, i.e., both odd and even states give the same
maximal BI violation $S_+$.
 This $S_+$ is maximally possible, $S_+=S_{\rm max}$, as shown by our
numerical calculations, so that
 \bes{4.14}
 \be
S_{\rm max}=2\lambda_a\sqrt{\max\left\{2,
\frac{\lambda_a^2+2\gamma_a-1}{\lambda_a^2-(1-\gamma_a)^2} \right\}}\
\ \mbox{if } \lambda_a^2>1-\gamma_a,
 \e{4.14a}
 \be
S_+=2\quad\mbox{if } \lambda_a^2\le1-\gamma_a.
 \e{4.14b}
 \ese
Now the optimal value of $\beta$ follows from Eq. \rqn{4.17} or
\rqn{5.26} to be
 \be
\beta=\pi/2-\arccos\{(\gamma_a-\gamma_a^2)/
[\lambda_a^2-(1-\gamma_a)^2]\}/2.
 \e{4.15}

Since qubit $b$ is not affected by decoherence, $S_+$ is invariant
with respect to arbitrary rotations of qubit $b$.
 Therefore, all states obtained from the optimal odd (or even) state
by rotations of qubit $b$ are optimal.
 [This explains why the states \rqn{4.16} and \rqn{5.12} yield the same
results: these states are related by the unitary transformation
$\sigma_x^b$ of the qubit $b$.]
 Moreover, since maximally entangled states transform to each other
by a rotation of one qubit \cite{vol00} (see Sec. \ref{IIIF}\,b),
$S_+$ is the same for all maximally entangled states [cf. Eq.
\rqn{5.19} or \rqn{5.20}],
 \be
 S_+=\left\{
 \begin{array}{ll}
 2\sqrt{2}\lambda_a,\ &\ T_1^a\le T_2^a, \\
 2\sqrt{\lambda_a^2+\gamma_a^2},&\ T_1^a\ge T_2^a.
 \end{array}\right.
 \e{4.20}

\begin{figure}[htb]
\includegraphics[width=8.cm]{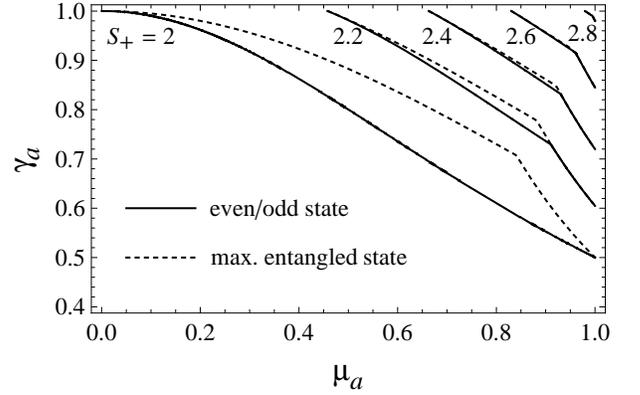}
\caption{Contour plot of the maximum $S_+$ of $S$ versus $\gamma_a$
and $\mu_a$.
  Solid lines: $S_+=S_{\rm max}$ \rqn{4.14} for the even state \rqn{5.12}
and the odd state \rqn{4.16}, dashed lines: Eq. \rqn{4.20} for a
maximally entangled state.}
 \label{f5}\end{figure}

Figure \ref{f5} shows the contour plot of $S_+$ versus $\gamma_a$ and
$\mu_a$ for the even and odd states which produce $S_+=S_{\rm max}$
(the solid lines) and for any maximally entangled state (the dashed
lines).
 Equation \rqn{4.14} and Fig. \ref{f5} imply that now the BI violation
is possible for $\gamma_a>0.5$, this limit being approached in the
absence of proper dephasing ($\mu_a=1$).
 Hence, in particular, for $\mu_a=1$ the BI violation duration
maximized over all states is given by
$\tau_B=T_1\ln2\approx0.693T_1$.
 Equations \rqn{4.14a} and \rqn{4.20} imply that the BI violation for
maximally entangled states is maximal ($S_+=S_{\rm max}$) when there
is no energy relaxation ($\gamma_a=1$) or when pure dephasing is
sufficiently weak, so that the the first term in the braces in Eq.
\rqn{4.20} is dominating (see Fig. \ref{f5}).
 Note that maximally entangled states produce practically the same
BI violation as the optimal states for $S_+\ge2.4$.

\subsubsection{General case}

In the general case when the decoherence parameters for the two
qubits are different, the maximum of the BI violation can be obtained
in a state of the form \rqn{5.7}, as discussed in Sec. \ref{IVC2}.
 We performed several hundred calculations of $S_+$ for the states
\rqn{5.7} and \rqn{5.12} with random values of the four parameters
$\gamma_k$ and $\mu_k\ (k=a,b)$ from the interval $[0.8,1]$.
 For this parameter range the values of $S_{\rm max}$ are greater
than 2.
 In our calculations $S_{\rm max}$ resulted from the
even state \rqn{5.12} in about 70\% of the cases.
 In the cases, where the even state did not yield the maximal $S$,
there were various optimal states, which included both general
maximally entangled states \rqn{3.27} and nonmaximally entangled
states.
 However, the difference between $S_{\rm max}$ and $S_+$ due to the
optimal even state was less than 0.1.

Thus, there can be several approaches of a various degree of
complexity and accuracy in order to obtain $S_{\rm max}$ and the
optimal observation conditions for given decoherence parameters:
 (a) One can use the exact numerical approach of Sec. \ref{IVC2},
which provides $S_{\rm max}$, the optimal state, and the optimal
detector configurations.
 (b) A simpler approach is to consider only the optimal even
state [Eqs. \rqn{5.16} and \rqn{5.26}], which provides rather
accurate, if not exact, result, as discussed above.
 (c) The analytical formula \rqn{5.13} and \rqn{4.17} for the optimal
odd state can be used, if, e.g., in the experiment the odd state is
realized more conveniently than other entangled states, as is the
case for experiments with superconducting phase qubits.
 (d) An even simpler approach is to use the Bell state
$|\Psi_+\rangle$ or $|\Phi_+\rangle$ [see Eqs. \rqn{5.19} and
\rqn{5.20}].
 In the approaches (c) and (d) the BI violation is obtained
under more restrictive conditions than in (a) and (b).
 However, if one requires a significant degree of the BI violation,
say, $S_+\ge2.4$ (which may be needed in the presence of other
experimental errors), there is practically no advantages for one
approach over the others (see Figs. \ref{f4} and \ref{f5}).

\section{Decoherence and measurement errors}
\label{V}

In the previous sections we assumed that measurements are ideal.
 Here we take into account the possibility that measurements of the
qubits are performed with local (independent) errors.
 Effects of local errors were studied elsewhere \cite{kof08,ebe93}.
 In this section we discuss combined effects of local errors and
local decoherence.
 The present case is rather involved, since it includes complications
due to both decoherence and errors.
 Here we discuss only the general approach to the problem, whereas
a detailed analysis is out of the scope of the present paper (see
also Ref. \onlinecite{kof08}).

\subsection{Description of errors}

In the frame of the error model considered in Ref.
\onlinecite{kof08}, the measured probabilities are written in the
form
 \be
p_{ij}^M=\sum_{m,n=0}^1F^a_{im}F^b_{jn}p_{mn} ={\rm Tr}(Q_i^a
Q_j^b\rho),
 \e{3.1}
where $Q_i^k=\sum_{m=0}^1F_{im}^kP_m^k\ (i=0,1;\ k=a,b)$ are the POVM
operators describing the measurement \cite{nie00} [cf. Eq. \rqn{13}]
and $F_{im}^k$ is the probability to find qubit $k$ in state
$|i\rangle$ when it is actually in state $|m\rangle$.
 The fidelity matrix $\{F_{im}^k\}$ obeys $F_{0m}^k+F_{1m}^k=1$ and
hence has two independent parameters, e.g., $F_0^k=F_{00}^k$ and
$F_1^k=F_{11}^k$, the measurement fidelities for the states
$|0\rangle$ and $|1\rangle$, respectively.
 Equation \rqn{3.1} implies that in the presence of local measurement
errors
 \be
S={\rm Tr}(\tilde{\cal B}\rho),
 \e{5.32}
where the error modified Bell operator \cite{kof08}
 \be
\tilde{\cal B}=\tilde{A}\tilde{B}-\tilde{A}\tilde{B}'+
\tilde{A}'\tilde{B}+\tilde{A}'\tilde{B}'.
 \e{5.18}
Here, e.g., $\tilde{A}=Q_1^a(\vec{a})- Q_0^a(\vec{a})$ [cf. Eq.
\rqn{3.24}] or
$\tilde{A}=\xi_-^a+\xi_+^a\vec{a}\cdot\vec{\sigma}_a$ and
$\tilde{B}=\xi_-^b+\xi_+^b\vec{b}\cdot\vec{\sigma}_b$, where
 \be
\xi_+^k=F_0^k+F_1^k-1,\quad\xi_-^k=F_1^k-F_0^k,
 \e{5.31}
$\tilde{A}'$ and $\tilde{B}'$ following from $\tilde{A}$ and
$\tilde{B}$ on the replacement of $\vec{a}$ and $\vec{b}$ by
$\vec{a}'$ and $\vec{b}'$, respectively.

\subsection{Modified Bell operator for decoherence and errors}

In the presence of decoherence and errors we perform the substitution
\rqn{5.1} in Eq. \rqn{5.32}.
 It is useful to recast the resulting expression as
 \be
S={\rm Tr}(\hat{\cal B}\rho_0)
 \e{5.30}
where the Bell operator modified by errors and decoherence is
 \be
\hat{\cal B}=({\cal L}_a^*\otimes{\cal L}_b^*)(\tilde{\cal
B})=\hat{A}\hat{B}-\hat{A}\hat{B}'+
\hat{A}'\hat{B}+\hat{A}'\hat{B}'.
 \e{5.27}
Here $\hat{A}=\hat{Q}_1^a(\vec{a})-\hat{Q}_0^a(\vec{a})$,
$\hat{B}=\hat{Q}_1^b(\vec{b})-\hat{Q}_0^b(\vec{b})$, etc., where the
POVM operators $\hat{Q}_i^k$ are given by $\hat{Q}_i^k={\cal
L}_k^*(Q_i^k)\ (k=a,b)$ or [cf. Eq. \rqn{5.4}]
 \be
\hat{Q}_i^k=\sum_{j=1}^3K_j^kQ_i^k(K_j^k)^\dagger
=\sum_{m=0}^1\sum_{j=1}^3F_{im}K_j^kP_m^k(K_j^k)^\dagger,
 \e{5.28}
so that [cf. Eq. \rqn{5.5}]
 \be
\hat{A}=\xi_-^a+\xi_+^a(\gamma_a-1)a_z+\xi_+^a\vec{q}_a
\cdot\vec{\sigma}_a,
 \e{6.1}
whereas the operators $\hat{A}'$, $\hat{B}$, and $\hat{B}'$ are given
by Eq. \rqn{6.1}, where $\vec{a}$ is replaced by $\vec{a}'$,
$\vec{b}$, and $\vec{b}'$, respectively.

Consider properties of $S$ which can be helpful in calculations.
 One of the peculiarities present in the case with errors is that the
maximal and minimal values of $S$ for a given state are generally not
equal by the magnitude \cite{kof08}, $S_+\ne |S_-|$.
 Note, however, the relations which follow from Eqs.
\rqn{5.30}, \rqn{5.27}, and \rqn{6.1},
      \bes{trSF}
    \bea
&& S \rightarrow -S \,\,\,\, \mbox{if} \,\,\,\, \vec{a}\rightarrow
-\vec{a},\,
  \vec{a}'\rightarrow -\vec{a}', \, F_0^a\leftrightarrow F_1^a ;
  \qquad
    \label{trSF-a} \\
&& S \rightarrow -S \,\,\,\, \mbox{if} \,\,\,\, \vec{b}\rightarrow
-\vec{b},\,
  \vec{b}'\rightarrow -\vec{b}', \, F_0^b\leftrightarrow F_1^b.
  \ea{trSF-b}
 \ese
Moreover, one has $S_+ = |S_-|$ if the two measurement fidelities are
equal, at least, for one qubit:
 \be
F_0^a=F_1^a\ \,\,\, \mbox{or} \,\,\, F_0^b=F_1^b .
 \e{4.23}
The relations \rqn{trSF} and \rqn{4.23} were obtained previously
\cite{kof08} for the case without decoherence.

 In the special case of equal measurement fidelities for each qubit,
 \be
F_0^a=F_1^a=F_a,\ F_0^b=F_1^b=F_b,
 \e{4.12}
Eq. \rqn{5.27} yields $\hat{\cal B}=(2F_a-1)(2F_b-1){\cal B}_d$ and
hence
 \be
S=(2F_a-1)(2F_b-1)S_d,
 \e{6.5}
where $S_d$ is the value of $S$ obtained in the presence of
decoherence but in the absence of measurement errors.
 This case can be analyzed, as discussed in Sec. \ref{IV}.

\subsection{Discussion}

There is no analytical solution in the presence of errors
\cite{kof08,ebe93}.
 Moreover, the present case involves an eight-dimensional parameter
space (there are two decoherence parameters and two measurement
fidelities for each qubit), which additionally complicates the
analysis.

Similarly to Sec. \ref{IV}, the modified Bell operator $\hat{\cal B}$
\rqn{5.27} can be used for numerical calculations, since for given
decoherence and error parameters the maximum (minimum) value of $S$
equals the maximum (minimum) of the greatest (smallest) eigenvalue of
$\hat{\cal B}$ over the detector directions, the optimal state being
given by the corresponding eigenvector.
 Now, as in Sec. \ref{IV}, $S$ is invariant only to rotations of the
qubits and detectors around the $z$ axis, which allows one to reduce
the number of the fitting parameters from eight to six by setting,
say, $a_y=b_y=0$.
 This number cannot be further reduced, since, in contrast to Sec.
\ref{IV}, in the presence of errors the set of optimal configurations
for a given state is not continuous \cite{kof08}, and hence it is
impossible to choose one of the optimal measurement axes at will.

This computation procedure is relatively slow.
 It produces generally different optimal states for different values
of the parameters.
 An approach which is faster and more relevant for most
experiments is to consider the BI violation for a specific initial
state $\rho_0$, e.g., the odd or even state [Eqs.  \rqn{4.16} and
\rqn{5.12}].
 In this case an expression for $S$ resulting from Eq. \rqn{5.30} is
varied over the detector directions and perhaps the state parameters
[e.g., $\beta$ in Eqs. \rqn{4.16} and \rqn{5.12}].

Note that in the cases of the odd and even states the number of the
detector parameters can be reduced from eight to seven.
 Indeed, since the odd (even) state is invariant under a rotation of
qubits $a$ and $b$ around the $z$ axis by an arbitrary angle $\alpha$
($\alpha$ and $-\alpha$, respectively), one can set, e.g., $a_y=0$,
in view of the $z$-rotation symmetry for $S$.
 For a detailed analysis of the case of the odd state, Eq.  \rqn{4.16},
see Ref. \onlinecite{kof08}.

\section{Conclusions}
 \label{VI}

In the present paper we have considered conditions for maximal
violations of the Bell inequality in the presence of decoherence.
 In addition, combined effects of decoherence and local measurement
errors have been discussed.

Since decoherence transforms a pure entangled state into a mixed
state, we have begun the consideration from the study of optimal
conditions for the violation of the BI \rqn{3} for a general (pure or
mixed) state.
 We have obtained all detector configurations providing the maximal
value of the parameter $S$ in Eq. \rqn{3} for an arbitrary state.
 We have shown that generally the set of all optimal configurations
for a given state is characterized by one continuous and one discrete
parameters, whereas in special cases it can be characterized by two
or three continuous parameters.
 We have obtained also the symmetry relation for the optimal detector
orientations, Eq. \rqn{3.32} or \rqn{3.44}, which follows from the
invariance of $S$ with respect to the qubit swap.

Further, we have considered effects of local decoherence on the BI
violation.
 The decoherence model used includes energy relaxation at the zero
temperature and arbitrary dephasing.
 We have employed the decoherence superoperator in the operator-sum
form with the Kraus operators which generalize the previously known
ones.
 We have expressed $S$ as the average over the initial state of
the Bell operator modified by decoherence.
 This operator has been used for numerical calculations in order to
obtain the maximal BI violation for any values of the decoherence
parameters.
 We have also developed a faster numerical approach, which is based
on an analytical solution for a certain class of states.

 We have studied the BI violation maximized over the observation
directions and over all states or a class of states and obtained all
corresponding optimal detector configurations.
 We have obtained simple analytical solutions for the odd and even
states [Eqs. \rqn{4.16} and \rqn{5.12} or, more generally, \rqn{4.7}
and \rqn{4.10}], which are often used in experiments on the BI
violation.
 In particular, the odd state is relevant for
experiments with superconducting phase qubits.
 Whereas in the absence of decoherence the optimal detector
configurations for the odd and even states are vertical (Sec.
\ref{IIIG3}), in the presence of decoherence the optimal
configurations are either vertical or horizontal.
 We have discussed both the general case of arbitrary decoherence
parameters and a number of important special cases.
 In particular, we have discovered that the even state is
optimal in most cases.
 Our analysis have been illustrated by numerical calculations.

We have considered the combined effects of local errors and
decoherence.
 In this case the maximal Bell violation depends on eight parameters.
 We have derived the Bell operator modified by decoherence and errors
and have used it to discuss symmetry properties of $S$ and special
cases.
 We have also outlined the numerical approaches in this case.

We have discussed the application of the above analysis to
superconducting phase qubits.
 However, the present results are applicable to many types of
qubits.
 Moreover, the present results have relevance to the ongoing
discussion of effects of decoherence on entanglement, a major
resource in the field of quantum information.

\acknowledgments
 I am grateful to A. N. Korotkov for useful
discussions and for the encouragement of this work.
 The work was supported by NSA and DTO under ARO grant
W911NF-04-1-0204.


\begin{thebibliography}{99}
\bibitem{note1}
An expression equivalent to Eq. \protect\rqn{3.26} but written in a
different basis was cited in Ref. \protect\onlinecite{vol00}.
\bibitem{note3}
Furthemore, one can make $\rho_{23}$ and $\rho_{14}$ nonnegative in
Eq. \rqn{3.69} by a local transformation.
 For instance, the $\pi$ rotation of one of the qubits around the
$z$ axis changes the signs of $\rho_{23}$ and $\rho_{14}$, whereas
the $\pi/2$ $z$-rotations of the qubits change the sign of
$\rho_{14}$.
\bibitem{note}
Here the symbols $\vec{a}$ and $\vec{a}'$ are interchanged in
comparison with Ref.\ \protect\onlinecite{hor95} in view of the
difference of the definitions of $S$ in the present paper and in
Ref.\ \protect\onlinecite{hor95} (cf. the last paragraph in Sec.
\ref{IIA}).
\bibitem{note2}
For the odd state, Eq. \rqn{5.15} is a special case for the zero
temperature of the result in Ref. \onlinecite{kof}.
\end{thebibliography}
\end{document}